\documentstyle[12pt,epsf]{article}

\newcommand{\la}[1]{\label{#1}}
\newlength{\templength}

\newlength{\numlen}
\newcommand{\n}{\settowidth{\numlen}{0}\makebox[\numlen]{}}
\newcommand{\cen}[1]{\multicolumn{1}{c}{#1}}

\newlength{\indexlength}

\newcommand{\be}{\begin{equation}}
\newcommand{\ee}{\end{equation}}
\newcommand{\ba}{\begin{eqnarray}}
\newcommand{\ea}{\end{eqnarray}}

\newcommand{\ie}{{i.e.},}

\newcommand{\eq}{eq.~}
\newcommand{\eqs}{eqs.~}

\newcommand{\rx}{\rangle}
\newcommand{\lx}{\langle}
\newcommand{\ff}{\widetilde}
\newcommand{\nr}[1]{(\ref{#1})}
\newcommand{\h}{{\hspace{0.5 cm}}}

\newcommand{\dd}{\mbox{d}}

\newcommand{\mod}{{\rm\,mod\,}}
\newcommand{\fr}[2]{{\frac{#1}{#2}}}
\newcommand{\per}[1]{\frac{1}{#1}}
\newcommand{\half}{{\scriptstyle{1\over2}}}

\newcommand{\Z}{{\bf Z}}    % defines integers
\newcommand{\R}{{\bf R}}    % real numbers
\newcommand{\3}[1]{{\bf #1}} % vector symbol
\newcommand{\lf}{{\rm L}}   % laboratory frame
\newcommand{\cm}{{\rm CM}}  % center of mass frame
\newcommand{\g}{\vec\gamma\,}
\newcommand{\Y}{{\cal Y}}   % defines the harmonic polynomial
\newcommand{\e}{{\bf e}}    % defines unit vector
\newcommand{\m}{m_{\phi}}    % light mass
\newcommand{\M}{m_{\rho}}   % heavy mass
\newcommand{\G}{\Gamma_{\rho}}
\newcommand{\kp}{\kappa_\phi}
\newcommand{\kr}{\kappa_\rho}
\newcommand{\kx}{g}

\begin{document}
% Title
\begin{titlepage}

\hfill IUHET 297

\hfill March 1995

\begin{centering}
\vfill

{RESONANCE SCATTERING PHASE SHIFTS ON A NON--REST FRAME LATTICE}

\vspace{1cm}
K.~Rummukainen
and
Steven~Gottlieb

\vspace{1cm}

{\em Department of Physics, Indiana University,\\
Bloomington, IN 47405, USA}

\vspace{2cm}
{\bf Abstract}

\end{centering}

\vspace{0.3cm}\noindent
Many low energy hadrons, such as the rho, can observed as resonances
in scattering experiments.  A proposal by L\"uscher enables one to
determine infinite volume elastic scattering phases from the
two-particle energy spectrum measured from finite periodic lattices.
In this work, we generalize the formalism to the case where the total
momentum of the particles is non-zero; \ie the lattice frame is not
the center-of-mass frame of the scattering particles.  There are
several advantages to this procedure including making a wider variety
of center of mass energies accessible with a fixed lattice volume, and
making the avoided level crossing in a P-wave decay occur with a
smaller volume.

The formalism is tested with a simple lattice model of two fields with
different masses and a 3-point coupling in $3+1$ dimensions.  We find
remarkable agreement between the rest-frame and non-rest-frame
scattering.

\vfill \vfill
\noindent
IUHET 297\\
\end{titlepage}

\section{Introduction\la{sec:intro}}

Numerical lattice simulations have been fairly successful in
calculating many low energy properties of QCD, the most fundamental
being the mass spectrum of the lightest
hadrons~\cite{SpectrumReviews}.  However, most hadronic states are
resonances that cannot be fully described without scattering with
stable asymptotic states.  A well-known example of this is the
$\rho$-meson, which appears as a resonance in the elastic
$\pi\pi\rightarrow\pi\pi$ scattering in the angular momentum $l=1$ and
isospin $I=1$ channel.  In fact, the calculation of the mass ratio of
the nucleon and $\rho$ has been a persistent problem for lattice
calculations and only recently has there been evidence that the
correct ratio emerges in the continuum
limit~\cite{Weingarten,Gottlieb94}.  This has been demonstrated in the
valence or quenched approximation where the $\rho$ cannot decay.  Most
current dynamical quark calculations are done in a regime where the
pions are so heavy that the rho cannot decay, so it is not that likely
that this effect has played in an important role in the dynamical
quark calculations~\cite{Altmeyer}; however, as the chiral limit is
approached this may become an important practical matter.

%%%%
Although hadronic coupling constants, which require calculation of
3-point functions, have been studied on the lattice quite some time
ago~\cite{Gottlieb86}, only recently have there been attempts to
resolve hadronic scattering phases, which require 4-point functions.
Most of the results are for meson--meson and meson--nucleon scattering
lengths at zero relative momentum [6--10]\@.  In the resonance channel
$\pi\pi$--$\rho$ the two pions are in the $l=1$ state and have
non-vanishing relative momentum.  The resonance has not yet been
observed on the lattice, despite an attempt by the MILC
collaboration~\cite{Bernard93}.

Elastic scattering involves at least four asymptotic states, which, in
the case of QCD, are themselves bound states of fundamental quark and
gluon fields.  Because the lattice simulations are by necessity
restricted to quite small and usually periodic physical volumes, the
direct observation of freely propagating asymptotic states is not
possible.  At best, the linear dimensions of the lattices are only a
few times the interaction length.

In a series of papers L\"uscher
\cite{Luscher86,Luscher91,Luscher91b} developed a method
that enables one to measure the scattering phase shifts from standard
lattice Monte Carlo simulations.  The method relates the two-particle
energy eigenstates on a finite periodic box to the infinite-volume
asymptotic scattering states.  The lattice measurement requires the
determination of the two-particle energy spectrum, a task readily
performed with lattice Monte Carlo methods.  The finite volume of the
lattice is not a disadvantage in this method, on the contrary, it is
used to probe the system, and simulations with several different
volumes are combined in order to obtain maximal knowledge about the
phase shift.%%%% new
\footnote{Related proposals for measuring the resonance
parameters on a finite lattice have been brought forward by Michael
\cite{Michael89}, Wiese \cite{Wiese89} and DeGrand \cite{DeGrand91}.}
In two dimensions, the method has been applied to O(3) nonlinear
$\sigma$-model \cite{Luscher90} and to a model consisting of two
coupled Ising spins \cite{Gattringer93}; it has also been applied to
3-dimensional QED \cite{Fiebig93,Fiebig94} and to 4-dimensional O(4)
$\phi^4$ theory \cite{Gockeler94}.
%%%% new
First attempts to apply the method to fermion-fermion scattering in
the Gross-Neveu model have been made \cite{Gockeler94b}.

In L\"uscher's formalism the center of mass of the scattering
particles is stationary with respect to the lattice, \ie the two
particles have opposite lattice frame momenta.  In this paper we
generalize the formalism to encompass elastic scattering with {\em
non-zero total momentum}.  This has several advantages: first, since
both the zero-momentum and non-zero momentum sectors can be measured
simultaneously, one lattice run may yield two separate data points
with a small cost in computer time.  (How small or large that cost may
be is model dependent and depends the ratio of update time to
measurement time.)  Second, the zero-momentum and non-zero momentum
data tend to complement each other: a much smaller volume range is
needed in order to obtain a good description of the phase shift as a
function of the center of mass momentum.  Third, as can be seen in
figure 1, for a decay where the outgoing particles must have relative
angular momentum as in $\rho\rightarrow\pi\pi$, the avoided level
crossing in the non-zero momentum sector occurs for a smaller lattice
size ($L/a\approx 15$ as opposed to $L/a\approx 35$, for the
parameters chosen in the figure).  Fourth, for channels that require a
vacuum subtraction in the zero momentum sector, the non-zero momentum
sector tends to have smaller statistical errors, since no subtraction
is required.

We apply the method to a simple 4-dimensional test model.  The model
contains two scalar fields: a light mass field $\phi$ and a heavier
field $\rho$.  Our goal is to investigate the scattering amplitude of
the infinite volume elastic $\phi\phi\rightarrow\phi\phi$ scattering
process.  The mass of the $\phi$ field cannot be too small: in order
to reduce the finite volume polarization effects (self-interactions of
the particles ``around the world'') the linear dimension $L$ of the
lattice should be at least a few times the correlation length $\xi =
1/\m$.  The field $\rho$ has a mass $\M\le 2\m$; with the introduction
of the 3-point $\rho\phi^2$ coupling the $\rho$ particle is rendered
unstable and it appears as a resonance in the $l=0$ scattering
channel.  Our model is simpler than nature where a charged pion field
has a derivative coupling to the spin-1 rho, $\pi^2\partial_\mu\pi
\rho^\mu$, resulting in a resonance in the $l=1$ channel.

In the center of mass system, the scattering amplitude $T$ of the
elastic scattering of two identical spin-0 bosons is a function of the
absolute value of the momenta of the incoming particles $p$ and the
scattering angle $\theta$.  The standard form of the partial wave
decomposition of the scattering amplitude is
\be
  T(p,\theta) = \fr{16\pi W}{p} \sum_{l=0}^{\infty} (2l+1)
		P_l(\cos\theta)t_l(p)\,,
\la{amplitude}
\ee
where $W = 2 \sqrt{p^2 + \m^2}$ is the total energy of the scattering
particles.  Assuming a unitary $S$-matrix the partial amplitudes
$t_l(p)$ are expressed with the phase shift $\delta_l(p) \in \R$ as
\be
  t_l(p) = \fr{1}{2i} (e^{i2\delta_l} - 1).
\la{partial-amplitude}
\ee
Due to the Bose symmetry $t_l$ vanish for odd $l$.

In this work we are primarily interested only in the $l=0$ channel,
where the resonance appears.  We measure the scattering phase shift
both in the zero and non-zero total momentum sectors and find an
excellent agreement between the momentum sectors and with the
perturbative ansatz.

The rest of this paper is organized as follows: in
sec.~\ref{sec:torus} we discuss the general properties of two-particle
states on a torus for free and then interacting particles.  In
sec.~\ref{sec:montecarlo}, we give the details of the Monte Carlo
simulations of our model, including the algorithms and parameters
used, the correlators calculated on the configurations, the energy
spectrum of various one- and two-particle levels, and the extraction
of the phase shift and resonance parameters.  In sections
\ref{sec:phaseshift} and \ref{sec:symmetry} we discuss the theoretical
aspects of the method: section \ref{sec:phaseshift} extends
L\"uscher's formalism to the non-rest frame case and derives the
relationship for the phase shift, eq.~\nr{deltal}, and
sec.~\ref{sec:symmetry} details the symmetry considerations that may
be used to simplify the wave functions for the energy
eigenstates. Brief conclusions are presented in
sec.~\ref{sec:conclusions}.

\section{Two-particle states on a torus\la{sec:torus}}

In this section we introduce the formalism necessary for calculating
the scattering phase shifts in a periodic 3-dimensional box.  Although
at the end the results will be applied to discrete periodic lattices,
at this stage we assume continuous space-time; and correct for
discreteness of the lattice structure later.  The formulae given here
are sufficient for analyzing the lattice data.  The derivation of the
basic equation~\nr{phil} is done in section \ref{sec:phaseshift}; the
reading of this section is not necessary for the analysis of the
lattice data.  We follow the basic formalism and notation introduced
by M.~L\"uscher~\cite{Luscher86,Luscher91}, generalizing it to
encompass the non-zero total momentum sector.

\subsection{Non-interacting particles\la{sec:noninteract}}

Let us first consider a system of two {\em non-interacting\,}
identical bosons ($\phi$) of mass $\m > 0$ in a cubic box of volume
$L^3$ with periodic boundary conditions.  The total energy of the
system in the rest frame of the box, which we shall call the
laboratory frame or the lattice frame (\lf) is
\be
W_\lf = \sqrt{\3p_1^2 + \m^2} + \sqrt{\3p_2^2 + \m^2},
\la{wlf}
\ee
where $\3p_i$ are the 3-momenta of the particles.  In the center of
mass frame (\cm) the energy is
\be
W_\cm = 2 \sqrt{p^{*2} + \m^2},
\la{wcm}
\ee
with $\3p^* = \3p^*_1 = -\3p^*_2$, $p^* = |\3p^*|$.  We denote here
the center of mass momenta with an asterisk $(^\ast)$.  In the
laboratory frame, the center of mass is moving with velocity $\3v$,
and the momenta $\3p_i$ and $\3p^*$ are related by the standard
Lorentz transformation
\be
\3p^* = \g(\3p_1 - \3v\sqrt{\3p_1^2 + \m^2}) =
       -\g(\3p_2 - \3v\sqrt{\3p_2^2 + \m^2})\,,
\la{lorentz}
\ee
where we have defined
\be
\gamma = \per{\sqrt{1-\3v^2}}
\la{defgamma}
\ee
and we have used the shorthand notation
\be
\g\3p = \gamma \3p_{\parallel} + \3p_{\perp}, \h
\g^{-1} \3p = \gamma^{-1} \3p_{\parallel} + \3p_{\perp},
\la{defg}
\ee
where $\3p_\parallel$ and $\3p_\perp$ are components of $\3p$ parallel
and perpendicular to the center of mass velocity: $\3p_\parallel =
(\3p\cdot\3v)\3v/v^2$ and $\3p_\perp =\3p -\3p_\parallel$.  Let the
laboratory frame total momentum $\3P = \3p_1 + \3p_2$; then, using
\eq\nr{lorentz}, the velocity of the center of mass can be written
as
\be
  \3v = \3P/W_\lf
\la{velocity}
\ee
and the center of mass and laboratory frame energies are related by
\be
W_\lf = \sqrt{\3P^2 + W_\cm^2} = \sqrt{\3P^2 + 4(p^{*2} + \m^2)}\,.
\la{wlfwcm}
\ee
Using \eq\nr{wlfwcm} we can obtain the center of mass momentum
$p^{*2}$ in terms of the laboratory frame momenta $\3p_i$:
\be
p^{*2} = \per4 (W_\lf^2 - \3P^2) - \m^2.
\la{plf}
\ee
It is illustrative to consider the energy levels of the system
consisting of two non-interacting particles.  In this case, the torus
quantizes the laboratory frame momenta $\3p_i$ to values
\be
  \3p_i=\fr{2\pi}{L}\,\3n, \h \3n\in\Z^3.
\la{freemom}
\ee
The total momentum is similarly quantized.  Note that the lowest
values of $\3p_i^2$ are evenly spaced: $p_i^2/(2\pi/L)^2 = 0,
1,2,\ldots$; the first skipped value is 7.  When the laboratory frame
total momentum $P = |\3P| = 0$, the momenta of the two particles are
opposite $\3p_2 = -\3p_1$ and the levels are given by
\eqs\nr{wlf} and \nr{freemom}:
\be
  W_\lf = W_\cm = 2\sqrt{\m^2 + (\3n 2\pi/L)^2},\h\h\3n\in\Z^3\,.
\la{freewlf}
\ee
In figure~\ref{fig:energylevels} we show the seven lowest $P=0$ levels
with solid lines as functions of $L$, where $L$ is given in units of
an arbitrary length scale $a$.  The mass of the particles is fixed to
$\m=0.3/a$.  The lowest level, corresponding to two particles at rest,
is the horizontal line at $W=0.6/a$.

\begin{figure}[tb]
\epsfxsize=14cm
\vspace{-1.2cm}
\centerline{\epsfbox{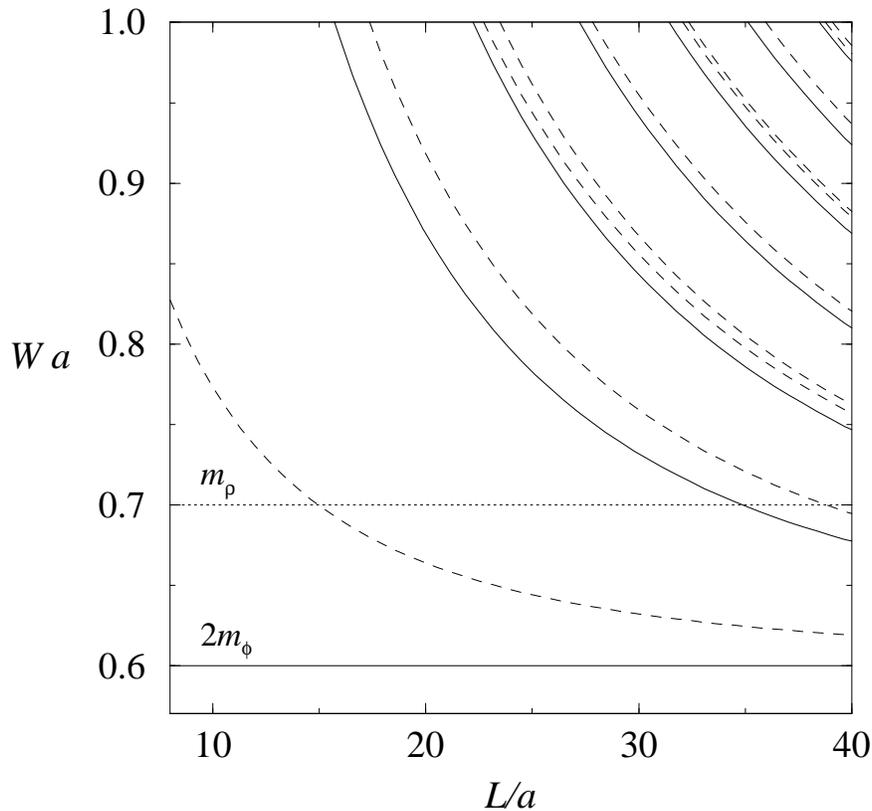}\hspace{1cm}}
\vspace{-7cm}
\caption[0]{Center of mass energy levels of a system of two
non-interacting particles of mass $0.3/a$ as functions of the system
size $L$.  Solid lines correspond to the zero total momentum sector
$P=0$, and dashed lines to sector $P=2\pi/L$.\la{fig:energylevels}}
\end{figure}

The lowest non-zero value for $P$ is $(2\pi/L)$ along some coordinate
axis.  For concreteness, let us choose $\3P=\e_3 2\pi/L$; now the
center of mass energy can be solved from \eq\nr{wlfwcm}:
\be
  W^2_\cm = W_\lf^2 - \3P^2
          = \left[\,\sqrt{\3p_1^2 + \m^2} +
		   \sqrt{\3p_2^2 + \m^2} \,\right]^2 - (2\pi/L)^2
\la{wlevels}
\ee
where $\3p_1 = (\3n+\e_3)\,2\pi/L$ and $\3p_2 = -\3n 2\pi/L$ for some
$\3n\in\Z^3$.  In figure~\ref{fig:energylevels} we show the lowest
center of mass energy levels with dashed lines; they correspond, from
the lowest upwards, to vectors
\be
  \3n = (0,0,0), (0,1,0), (0,0,1), (1,1,0), (0,1,1),
   (0,2,0), (1,1,1), (1,2,0), \ldots
\ee
where we have chosen $n_2\ge n_1\ge 0$; momentum states with
$(n_1,n_2)\rightarrow(\pm n_1,\pm n_2)$ or $(\pm n_2,\pm n_1)$ have
degenerate energy levels.

It should be noted that the center of mass frame is not unique,
because the velocity of the center of mass for each two-particle state
depends on the energy $W_\lf$ according to
\eq\nr{velocity}.
Thus, even for fixed $L$, we cannot make a boost to {\it the} center
of mass frame and measure the energy levels there.  The most
convenient choice is to perform all the measurements in the laboratory
frame, where all the levels have the same total momentum.  {}From
\eqs\nr{velocity} and \nr{wlevels} we note that the velocity of the
center of mass frame with respect to the laboratory frame is given by
$v=[1 + (W_\cm L/2\pi)^2]^{-1/2}$.

When comparing the sectors $P=0$ and $P=(2\pi/L)$ we note that for
each $P=0$ level there is a $P\ne0$ level with a slightly higher
energy.  The largest difference is between the lowest levels, where
the $\3p_1 = \3p_2 = 0$ level does not have any volume dependence.
Occasionally there appear two closely spaced levels in the
$P=(2\pi/L)$ sector.  This is a relativistic effect, that lifts the
degeneracy of some of the momentum states.  As an example, let us
consider two-particle states with momenta ($p_1=4\pi/L, p_2=2\pi/L$)
and ($p_1 = 2\sqrt{2}\,\pi/L$, $p_2 = 2\sqrt{3}\,\pi/L$).  These
states correspond to vectors $\3n = (0,0,1)$ and (1,1,0) in
\eq\nr{wlevels}, which gives them non-degenerate laboratory frame
energies, but both have non-relativistic kinetic energy
\be
  W_{\lf,\rm nonrel} = \fr{p_1^2
+ p_2^2}{2\m} = 5\fr{(2\pi/L)^2}{2\m}\,.
\ee
The next non-zero total momentum sector is $P = \sqrt{2}\,2\pi/L$,
corresponding, for example, to $\3P=(\e_1 + \e_2)2\pi/L$.  In this
case already the lowest two 2-particle energy levels are nearly
degenerate.

\subsection{Interacting particles}

The utility of the $P\ne 0$ momentum sectors on a lattice becomes
evident when we consider resonance scattering.  To illustrate this let
us introduce a new particle ($\rho$) with mass $\M$, $2\m<\M<4\m$, and
with the same laboratory frame momentum as the two $\phi$ particles.
In figure~\ref{fig:energylevels} the $\rho$ particle energy level is
shown with a horizontal dotted line $W=\M$; we have arbitrarily chosen
$\M=0.7/a$.  The $\rho$ energy level is intersected by $2\phi$ energy
levels at various system sizes $L$.  If we now introduce a small
3-point $\phi\phi-\rho$ interaction, $\rho$ becomes unstable.  Due to
the interaction the energy eigenstates are a mixture of $\rho$ and
$2\phi$ states.  This is evident as {\em avoided level crossings} in
the energy levels; in figure~\ref{fig:energylevels} the interaction is
absent, but the avoided crossings would take place near the
intersections of the $\rho$ and $2\phi$ levels.  For the $P=0$ sector
the first crossing occurs at $L=35\,a$, whereas for the $P=2\pi/L$
sector the first crossing is already at $L=15\,a$.  In numerical
lattice simulations this difference means that the avoided level
crossing can be observed on lattices with more than 10 times smaller
spatial volume.

In our calculations the $\phi^4$ interaction in action \nr{cont-act}
is strong enough so that the two-particle energy levels are not close to
the free particle levels even far away from the level crossings.  When
the interaction is turned on the momenta $\3p_i$ of the individual
particles are not good quantum numbers and the quantization condition
\nr{freemom} is not valid any more.  On the other hand, since the
interaction depends only on the relative position of the particles,
the total momentum is conserved and is still quantized by
\eq\nr{freemom}.  For compactness, we shall label the different
$\3P$-sectors with a vector $\3d\in\Z^3$, defined through
\be
  \3P = \3d\fr{2\pi}{L}\,.
\ee
Assuming that the interaction is localized to a region smaller than
$L$, the main result of section~\ref{sec:phaseshift} is that the
energy spectrum is still given by the formulae \nr{wlf}, \nr{wcm} and
\nr{wlfwcm} but instead of the the momentum quantization condition
\eq\nr{freemom} we have a relation involving the scattering phase
shifts $\delta_l$, where $l$ labels the angular momentum of the
scattering channel.  Assuming that the phase shifts $\delta_l$ with
$l=2$, 4, 6,\ldots are negligible in the energy range of interest,
the phase shift $\delta_0$ is related to the momentum $p^*$ by
\be
  \delta_0(p^*) = -\phi^\3d(q) \mod \pi, \h q = \frac{p^*L}{2\pi}
\la{deltal}
\ee
where $\phi^\3d$ is a continuous function defined by the equation
\be
  \tan (-\phi^\3d(q)) = \frac{\gamma q \pi^{3/2}}{Z^\3d_{00}(1;q^2)}
  \h\h\phi^\3d(0) = 0.
\la{phil}
\ee
Function $Z^\3d_{00}$ is generalized zeta function, and is formally
given by
\be
  Z^\3d_{00}(s;q^2) = \per{\sqrt{4\pi}}
		\sum_{\3r\in P_{\rm d}} (\3r^2 - q^2)^{-s}\,,
\la{zetaf00}
\ee
where the set $P_\3d$ is
\be
  P_\3d = \{\3r \in\R^3 |
		\3r = \g^{-1}(\3n + \3d/2),\,\,\3n \in \Z^3\}\,.
\ee
The expansion \nr{zetaf00} is convergent when ${\rm Re\,}s>3/2$, but
it can be analytically continued to $s=1$.  We discuss the numerical
evaluation of $Z^\3d_{00}$ in section
\ref{sec:evaluate-zeta}.

When we select the sector $\3d=0$, $\gamma=1$, the formulae
(\ref{phil}--\ref{zetaf00}) become identical to eqs.~(1.3--1.5) in
ref.~\cite{Luscher91}.  In this case, the channel $l=2$ decouples from
the $l=0$ -channel, and it is sufficient to assume that phase shifts
$\delta_4$, $\delta_6$,\ldots can be neglected.  Because we are
interested in the elastic two-particle scattering, the center of mass
energy is restricted to the interval $2\m < W_\cm < 4\m$, or
equivalently $0< p^* < \sqrt{3}\m$.

\begin{figure}[tb]
\epsfxsize=14cm
\vspace{-1.2cm}
\centerline{\epsfbox{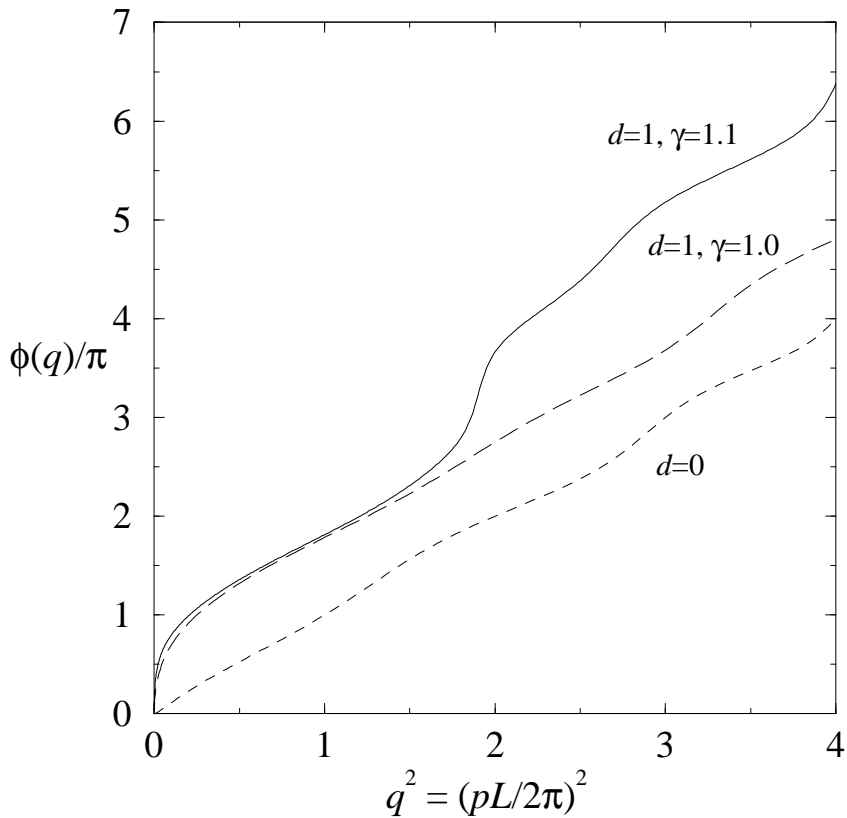}\hspace{1cm}}
\vspace{-7cm}
\caption[0]{Function $\phi^\3d(q)$.  Dashed line is $\phi$ when
the total momentum $\3P=0$ ($\3d=0$), line with long dashes with
$\3P=2\pi/L$ and $\gamma=0$, and solid line $\3P=2\pi/L$ and
$\gamma=1.1$\,.\la{fig:phi}}
\end{figure}

The function $\phi^\3d$ is a function of 3 variables, $q$, $\gamma$
and $\3d$.  In figure \ref{fig:phi} we display $\phi^\3d$ when
$d=|\3d|=0$; $d=1$ and $\gamma=1$; and $d=1$ and $\gamma=1.1$.  The
points where $\phi^\3d$ crosses levels $n\pi$ correspond to {\em free
particle\,} energy levels; when $d=0$ these occur at $q^2 =
0,1,2,3,\ldots$, in accordance with \eq\nr{freewlf}.  The curve
$d=1,\,\gamma=1$ is an artificial case and in effect neglects
relativistic effects: in this case the momenta transform according to
Galilean transformations, and the free particle energy levels are at
$q^2 = \fr14,\,\fr54,\,\fr94,\ldots$.  The third curve has $d=1$ and
$\gamma$ has an arbitrary value of 1.1.  As discussed above, the
relativistic effects lift the degeneracy of some momentum states; this
is evident by the rapid step-like cross-overs in the value of
$\phi^\3d$ by an amount of $\pi$ at some values of $q^2$.  When
$\gamma$ is decreased towards value 1 the breaking of the degeneracy
becomes smaller, and the cross-overs of $\phi^\3d$ approach a step
function.

\section{Monte Carlo simulations of the test model\la{sec:montecarlo}}

To observe a resonance, we construct a model with a light mass field
($\phi$) coupled to a field of heavier mass ($\rho$) with a 3-point
coupling.  The physical masses of the fields are in the elastic region
$2\m<\M<4\m$.  A computationally simple model with these features can
be constructed by coupling two Ising fields with different masses.  In
two dimensions, this model has been successfully used by Gattringer
and Lang to study the zero total momentum resonance scattering
\cite{Gattringer93}.  With this model it is straightforward to adjust
the masses of the particles and the strength of the coupling.

The Ising model is one representation of lattice $\phi^4$ models.
These models are most probably trivial and reduce to a free theory in
the continuum limit.  However, for small lattice spacings $a$ and at
energies substantially below the cutoff scale $1/a$ the model exhibits
a well-defined scaling behavior, corresponding to an interacting
continuum theory.  In the scaling region the behavior of the $\phi^4$
model is very well described with perturbative methods; this has been
studied in detail both analytically \cite{Luscher87} and with lattice
Monte Carlo methods \cite{Montvay87,Montvay88}.

\subsection{Continuum action and perturbative phase shift}

In Euclidean continuum formulation, the action of the model we
consider in this work is
\be
  S = \int\dd^4x \left[
	\fr12(\partial_\mu\phi)^2 + \fr12 m_{\pi,0}^2\phi^2 +
	\fr{\lambda_{0}}{4!} \phi^4 +
	\fr12(\partial_\mu\rho)^2  + \fr12 m_{\rho,0}^2\rho^2 +
	\fr{\lambda_{\rho,0}}{4!} \rho^4 +
	\fr{\kx_0}{2} \rho\phi^2 \right]\,.
\la{cont-act}
\ee
In practice, the coupling constant $\kx$ is small enough that we
expect the model to exhibit a scaling behavior similar to that of the
$\phi^4$ model.  The parameters of the model are chosen so that the
fields $\phi$ and $\rho$ have physical masses in the elastic region
$2\m\le\M<4\m$.

The resonance scattering $\phi\phi\leftrightarrow\rho$ occurs through
the $l=0$ channel.  The corresponding scattering phase shift
$\delta_0$ can be calculated perturbatively from action \nr{cont-act}.
At the resonance the partial amplitude becomes singular, and we can
approximate the total phase shift as a sum of the singular
contribution $\delta_s$ and the regular part $\delta_r$:
\ba
  \delta_0 &=& \delta_r + \delta_s  \la{deltasum} \\
  \delta_r &=& -\lambda_\1R \fr{p}{16\pi W} +
	\fr{\kx_\1R^2}{32\pi}\fr{1}{Wp}
	\log\fr{4p^2+\M^2}{\M^2} \la{regular} \\
  \tan\delta_s &=& -\fr{\kx_\1R^2}{16\pi}\fr{p}{W}
	\fr{1}{W^2-\M^2} \,. \la{singular}
\ea
The masses $\m$ and $\M$ are the physical masses of the particles,
obtained from the real part of the propagator poles.  The couplings
$\lambda_\1R$ and $\kx_\1R$ are renormalized $\phi^4$ and $\phi^2\rho$
coupling constants.  The term proportional to $\lambda_\1R$ in
\eq\nr{regular} arises from the 4 point diagram.  The $\kx_\1R^2$
terms come from the $\rho$ particle interchanges; in \eq\nr{regular}
from the sum of u and t channels and in \eq\nr{singular} from the s
channel.  The resonance width $\G$ is
\be
  \G = \fr{\kx_\1R^2}{32\pi\M^2}\sqrt{\M^2 - 4\m^2}\,.
\la{reswidth}
\ee
Besides the appearance of the $\lambda_\1R$ term and different
numerical factors, the formulae (\ref{deltasum}--\ref{singular}) are
similar to the perturbative results in O(N) symmetric $\phi^4$ theory
in ref.~\cite{Gockeler94}.

\subsection{Lattice action}

Let us consider a lattice of size $L^3\times T$ with periodic boundary
conditions.  For simplicity, we set the lattice spacing to unity, and
an appropriate power of the lattice spacing is understood in all
dimensionful quantities.  The model is defined by the action
\be
  S = -\kp \sum_{x;\,\hat\mu} \phi_x \phi_{x+\hat\mu}
      -\kr \sum_{x;\,\hat\mu} \rho_x \rho_{x+\hat\mu}
      +\kx \sum_{x;\,\hat\mu} \rho_x \phi_x \phi_{x+\hat\mu}\,.
\la{lattaction}
\ee
In the above summations $x$ goes over the whole lattice and the vector
$\hat\mu =\pm\e_\mu$, where $\e_\mu$ denotes a unit vector to
direction $\mu$.  The field variables $\phi$ and $\rho$ are restricted
to values $\{-1,+1\}$\@.  Since $\phi^2 =1$, the 3-point term has been
introduced in a nonlocal way.  The hopping parameters $\kp$ and $\kr$
are restricted to positive values, corresponding to ferromagnetic
Ising coupling.

If the coupling constant $\kx = 0$, the model corresponds to two
independent Ising models.  In infinite volume, the 4-dimensional Ising
model has a second order phase transition at $\kp=\kr\approx
0.074834(15)$ \cite{Gaunt79}.  When $\kx\ne0$, the third term in the
action \nr{lattaction} gives rise to the interaction
$\phi\phi\leftrightarrow\rho$, and by suitably choosing the couplings
$\kp$ and $\kr$ the $\rho$ particle is rendered unstable.  Since
$\rho$ is not an asymptotic state, its mass becomes ill-defined; in
what follows we mean by the $\rho$ mass $\M$ the energy level where
the phase shift reaches the resonance value.  Further, the action now
preserves the symmetry $\phi\rightarrow-\phi$ but the symmetry
$\rho\rightarrow-\rho$ is broken.  This has implications for the phase
structure of the model.  Let us discuss the case where all coupling
constants are greater than 0 but finite.  When $\kx$ is small enough,
the effective hopping parameter for the $\phi$ field becomes local,
but it remains still positive: $\kappa_{\rm eff}(x,\mu) =\kp-\half\kx
(\rho_x +\rho_{x+\hat\mu}) > 0$.  For a fixed $\rho$ configuration,
the action of the $\phi$ field corresponds to a ferromagnetic spin
glass, which still has an Ising-like second order phase transition
between phases where $\lx\phi\rx=0$ and $\lx\phi\rx\ne0$ in the
thermodynamic limit ($L,\,T\rightarrow\infty$).  It is not known to us
whether this transition survives the dynamic $\rho$ field.  However,
this seems plausible, since the expectation value of $\phi$ is still
driven to zero when $\kp$ is small enough (but still $> 0$) and
acquires a non-zero value with large enough $\kp$.

When $\kp > \kx$ the expectation value $\lx\phi_x\phi_{x+\hat\mu}\rx
>0$, and the third term of the action \nr{lattaction} acts like a
biased random magnetic field on the $\rho$-configuration.  The
expectation value of $\rho$ remains non-zero at all finite values of
the hopping parameter $\kr$, and hence the Ising phase transition of
the $\rho$ field disappears.  Thus, for any fixed $\kr$ there is
probably an upper limit on the (disconnected) correlation length of
the $\rho$ field, corresponding to a lower limit on $\M$ in lattice
units.  $\M$ can be made smaller only by simultaneously tuning $\kr$
and $\kx$.

The model described by the action \nr{lattaction} is most probably
trivial like the Ising model, and it does not have a continuum limit.
However, with small but finite lattice spacing the model effectively
describes an interacting continuum field theory.  For our purposes
this is sufficient, since our main interest is in the technical
methods for extracting the scattering phase shift from an interacting
effective theory.  However, this makes the analysis of the effects
caused by the finite lattice spacing by the standard method of
performing simulations with several different lattice spacings
difficult: models with different lattice spacings (correlation
lengths) correspond to different continuum theories.

\subsubsection{Update algorithm}

The spins are updated with a 4-dimensional analogue of the cluster
algorithm already used in 2-dimensional simulations
\cite{Gattringer93}.  The $\phi$ and $\rho$-spins are updated in separate
cluster update sweeps using modified Swendsen-Wang \cite{Swendsen87}
type percolation cluster algorithms.

In the $\phi$ update the clusters are grown using the local hopping
parameter $\kappa_{\rm eff}(x,\mu)=\kp-\half\kx(\rho_x
+\rho_{x+\hat\mu})$.  The link from point $x$ to point $x+\hat\mu$ is
{\em activated\,} with the probability $(1-e^{-2\kappa_{\rm
eff}(x,\mu)})\,\delta(\phi_x,\phi_{x+\hat\mu})$.  The clusters are
grown over the activated links with the Hoshen-Kopelman
algorithm~\cite{Hoshen76}.  After all the clusters are formed, all of
the spins in each cluster are set to value $+1$ or $-1$ with equal
probabilities.

In the update of the $\rho$ field the links are activated with
probability $(1-e^{-2\kr})\,\delta(\rho_x,\rho_{x+\hat\mu})$.  After
the clusters are formed, we calculate the net external field induced
by the $\phi$ field acting on the cluster $C$, and the spins in $C$
are set to value $+1$ with probability
\be
  p_+ = \Big(1+\exp[\kx \sum_{x\in C;\mu}
	\phi_x(\phi_{x+\hat\mu} + \phi_{x-\hat\mu})]\Big)^{-1}\,.
\ee
The probability of setting the spins in cluster $C$ to $-1$
is $p_- =1 - p_+$.

For our choices of the run parameters the cluster update is a clear
improvement over the local Metropolis algorithm, especially for the
$\phi$ field, which has longer correlation length.  For the $\rho$
field update the difference between the local and cluster update is
smaller, but still appreciable.

\subsection{The simulation parameters}

In order to achieve the resonating behavior the couplings in the
action \nr{lattaction} have to be tuned so that the $\rho$ resonance
lies in the elastic region: $2\m < \M < 4\m$.  For each (fixed)
coupling $\kx$, we adjust the hopping parameters $\kp$ and $\kr$ so
that the $\phi$ field is in the symmetric phase ($\lx\phi\rx=0$) and
the above inequalities hold.  The hopping parameters were tuned by
performing simulations with small lattices and measuring the single
particle correlation lengths for $\phi$ and $\rho$ fields (the
resonance turns out to be so narrow that the single particle
correlation length yields a very good estimate of the resonance mass
for practical lattice sizes).

The production runs were performed with 3 different sets of the
coupling constants, and the values of the couplings are listed in
table \ref{table:runs}.  The parameter sets are labelled with symbols
A, B and C\@.  In case A we set the coupling constant $\kx=0$ to test
the calculational methods in the absence of the 3 point coupling; B
and C have $\kx = 0.008$ and $0.021$, respectively.  Sets A and B
correspond roughly to masses $\m a\sim 0.2$ and $\M a\sim 0.5$;
however, no special attempt was made to tune the masses very
precisely.  In the case C we set $\m a\sim 0.3$ and $\M a\sim 0.8$ in
order to be able to use smaller lattice volumes and save computation
time.  Even in this case the finite lattice spacing effects turn out
to be well in control (sect.~\ref{sec:single}).

\begin{table}[tb]
\setlength{\templength}{\tabcolsep}
\setlength{\tabcolsep}{0.9cm}
\center
\begin{tabular}{l|lll}\hline
              &\cen{A}    &\cen{B}    &\cen{C}     \\ \hline\hline
$\kp$         & 0.0742    & 0.07325   & 0.07075    \\
$\kr$         & 0.0708    & 0.0718    & 0.0665     \\
$\kx$         & 0         & 0.008     & 0.021      \\
\# of lattice sizes & 9    & 10        & 16   \\ \hline
$\m\,a$       & 0.1856(4) & 0.1996(5) & 0.3081(4)  \\
$\M\,a$       & 0.5049(5) & 0.5306(13)& 0.8206(11) \\
$\G\,a$       & 0      & 0.0044(2) & 0.0178(7)  \\
$\kx_\1R \,a$ & 0         & 0.598(14) & 1.49(3)    \\
$\lambda_\1R$ & 28.1(1.1) & 36.8(1.3) & 48.3(2.0)  \\
%$Z_\pi $      &           &           &            \\
%$Z_\rho$      &           &           &            \\
\hline
\end{tabular}
\caption[1]{Compilation of the run parameters and results of the mass
and resonance parameter measurements.\la{table:runs}}
\setlength{\tabcolsep}{\templength}
\end{table}

The sets A, B and C consist of 9, 10 and 16 different lattice sizes,
respectively.  The size of the largest lattice in each of the sets is
$36^3\times 40$.  In all of the lattices the time extent $T$ was
larger than the spatial lattice length $L$.

The Monte Carlo simulations consisted of 30\,000 -- 65\,000 cluster
update sweeps for both of the spins for each lattice.  The correlation
function measurements, described in sections \ref{sec:single}
and~\ref{sec:two}, were performed after every 2 full update sweeps,
and 500 consecutive measurements were blocked together.  The final
error analysis was done by applying the jackknife method to the
blocked data.  At the end, we had 30 -- 65 jackknife blocks for each
lattice.  In what follows, the cited error bars are all only
statistical one standard deviation errors.  The data collected are
quite sufficient for measuring the correlation functions to the
accuracy needed in order to extract the phase shift.  Only for the
largest volumes does the statistical noise start to affect the signal
to any large extent.  The simulations were performed on IBM RS/6000,
DEC Alpha and Sun workstations, and the total CPU time used was
approximately 1800 hours.

\subsection{Single particle spectrum
and lattice effects\la{sec:single}}

In order to calculate the phase shift, we need an accurate
determination of the single particle mass of $\phi$ field $\m$.  The
single particle energy spectrum gives also an estimate of the finite
lattice spacing and finite volume effects.  These effects are expected
to be small provided $\m L \gg 1$ and $|\3p|\ll 1$.

In order to measure the correlations, we take the spatial Fourier
transform of the field $\phi$:
\be
  \ff\phi(\3n,t) = \fr1{L^3} \sum_{\3x} \phi(\3x,t) e^{i\3x\cdot\3p}
\la{fourier}
\ee
where $\3p = (2\pi/L)\3n$, and $-L/2+1\le n_i\le L/2$.  The single
particle energy spectrum can be extracted from the exponential decay
of the correlation functions
\be
  C_i(t) = \Big\langle \sum_{\3n;\,\3n^2 = i}
	     \ff\phi(-\3n,t) \ff\phi(\3n,0) \Big\rangle
 	     \propto e^{ -E(\3p) t}.
\la{corrfunct}
\ee
For each lattice, we measure 5 $\phi$ field correlation functions
$C_0$ -- $C_4$, corresponding to momenta $\3pL/(2\pi) = (0,0,0)$,
(1,0,0), (1,1,0), (1,1,1), (2,0,0) and permutations.

\begin{figure}[tb]
\epsfxsize=14cm
\vspace{-1.4cm}
\centerline{\epsfbox{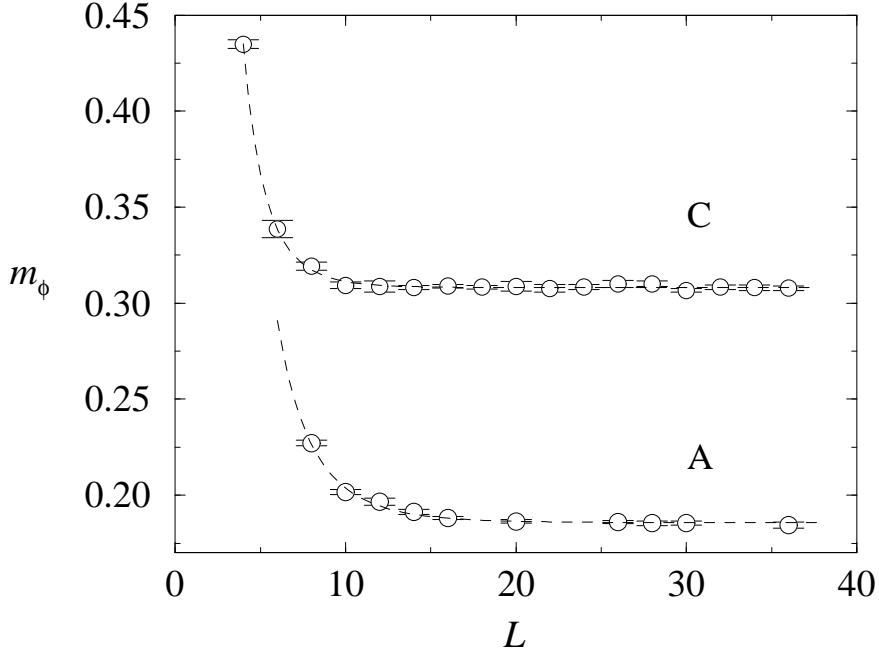}\hspace{1cm}}
\vspace{-9cm}
\caption[0]{The finite size dependence of the mass of the $\phi$
field for coupling constant sets A (bottom) and C (top).  The results
of the fits of function \nr{finite-size} are shown with dashed
lines.\la{fig:finite-size}}
\end{figure}

The single particle mass is extracted from the zero momentum
correlation function $C_0(t)\sim e^{-\m(L) t}$.  We expect that the
finite volume polarization -- the self-interactions of the field
around the periodic torus -- affects the mass as
\be
  \m(L) = \m + c\, L^{-3/2}\, e^{-\m L}\,,
\la{finite-size}
\ee
with a constant $c$.  The infinite volume mass $\m$ is obtained by
fitting the function \nr{finite-size} to the measured mass values
$\m(L)$.  The result of the fits are given in table~\ref{table:runs},
with one standard deviation errors.  In figure~\ref{fig:finite-size}
we show the data and the fitted function for cases A and C; case B is
very similar to case A.  In case C, the finite size effects
practically disappear into the statistical noise when $L\ge 12$.

\begin{figure}[tb]
\epsfxsize=14cm
\vspace{-1.5cm}
\centerline{\epsfbox{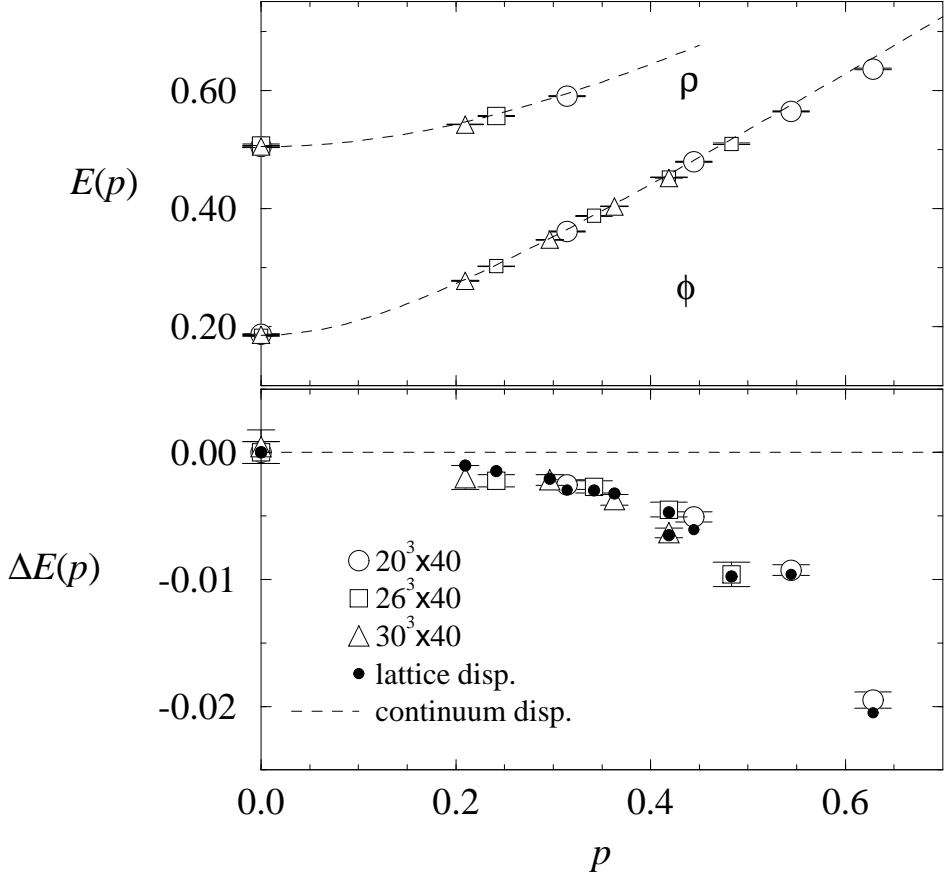}\hspace{1cm}}
\vspace{-6cm}
\caption[0]{The single particle energy levels of case A lattices
 of spatial size $20^3$, $26^3$ and $30^3$.  Top: the measured energy
levels together with the continuum dispersion relation,
\eq\nr{contdispersion} (dashed line) for $\phi$ and $\rho$.
Bottom: the measured $\phi$ energy levels and the lattice dispersion
relation \eq\nr{dispersion} (black dots); here the continuum
dispersion relation has been subtracted to emphasize the
difference.\la{fig:single}}
\end{figure}

In order to study the effects of the finite lattice spacing, we
compare the non zero momentum single particle energy levels to the
continuum relation
\be
  E(\3p) = \sqrt{\3p^2 + \m^2}
\la{contdispersion}
\ee
and to the lattice dispersion relation, valid for Gaussian fields on
the lattice
\be
  \cosh E(\3p) - \cosh\m = \sum_{i=1}^3 (1-\cos p_i).
\la{dispersion}
\ee
In figure~\ref{fig:single} we display the measured values of $E(p)$
and the continuum and lattice dispersion relations,
eqs.~\nr{contdispersion} and~\nr{dispersion}, for 3 lattices belonging
to series A.  The continuum dispersion relation is clearly
insufficient, and diverges from the measured dispersion $\sim p^2$.
On the other hand, the lattice dispersion relation fits the measured
energies strikingly well -- in the bottom part of
figure~\ref{fig:single} the difference between the data and the
lattice dispersion can be completely attributed to statistical errors.
The other lattice volumes and cases B and C behave in a qualitatively
similar fashion.  It should be noted that the lattice dispersion
relation~\nr{dispersion} is defined only at discrete momenta $p_i =
(2\pi/L)n_i$, $n_i\in\Z$.

In case A, the $\rho$ particle is stable, and we can check the finite
size and the finite lattice spacing effects on $\rho$ energy levels.
The mass spectrum is measured with analogous operators to
\eq\nr{fourier}, but we only measure momentum states $p=0$ and
$p=2\pi/L$.  In the zero momentum case, we do not find any appreciable
finite size effects, and the value of $\M$ given in
table~\ref{table:runs} is an average over all lattices.  The non zero
momentum energy levels are shown in figure~\ref{fig:single}.  They
also deviate from the continuum dispersion relation, but are well
described by the lattice dispersion.

The finite lattice spacing effects are large enough that we cannot use
the continuum dispersion relation when we determine the scattering
phase shifts.  However, the use of the lattice dispersion relation
\nr{dispersion} can be problematic, since the momenta in this
case are not restricted to ``lattice momenta'' $\3n 2\pi/L$ any more.
This is discussed further in section~\ref{sec:two}.

\subsection{Two-particle spectrum\la{sec:two}}

To measure the two-particle energy levels we need operators of type
\be
  \0O_{\3d,\3n}(t) = \ff\phi(\3n+\3d,t)\ff\phi(-\3n,t)
\la{2part-operator}
\ee
where $\ff\phi$ is the spatial Fourier transform of the $\phi$ field,
defined in \eq\nr{fourier}, and $\3d\,,\3n\in\Z^3$.  This operator
describes two $\phi$ particles with total 3-momentum $\3P =
(2\pi/L)\,\3d$.  For each value of vector $\3d$ the operators
\nr{2part-operator} form a complete basis for two $\phi$ particles.
The energy eigenstates can be constructed as linear combinations of
these operators and the $\rho$ operator belonging to the same total
momentum sector.  On an $L^3$ spatial lattice there are $L^3$
different operators of type \nr{2part-operator} for each value of
$\3d$; however, it turns out that only a few lowest momentum operators
are needed.  The energy eigenstates are measured as follows: first, we
pick a suitable set of operators $\0O _\alpha$, measure a matrix of
correlation functions $\lx\0O^{\dagger}_\alpha(t)
\0O_\beta(0)\rx$, and diagonalize the matrix.  The energy
eigenvalues can be extracted from the exponential decay of the
eigenvalues of the correlation function matrix.

To construct the operators \nr{2part-operator} we first form the
Fourier transforms of the field $\phi$ for momenta $\3pL/(2\pi) =
(0,0,0)$, (1,0,0), (1,1,0), (1,1,1), (2,0,0), and non-equivalent
permutations and signs, all in all 17 functions.  (Note that because
the field is real there is no need for separate computation of $-\3p$
and $\3p$.)  In the $\3P=0$ sector, these are combined into 5
different operators
\be
  O_{0,i}(t) = \fr{1}{N_i} \sum_{\3n} \delta_{\3n^2,i-1} \,
	\ff\phi(\3n,t)\ff\phi(-\3n,t)
	\h\h i=1,2,3,4,5\,,
\la{operator-0}
\ee
where $N_i = \sum_{\3n} \delta_{\3n^2,i-1}$.  In the $d=|\3d|=1$
sector we have 4 operators for each of the 3 directions of $\3d$.  For
concreteness, let $\3d=\e_3$; now the operators are
\ba
  O_{\e_3,1}(t) &=& \ff\phi(\e_3,t)\ff\phi(0,t)    \nonumber \\
  O_{\e_3,2}(t) &=& \fr14\sum_{\3n} \delta_{n_3,0}\,\delta_{n^2,1}\,
		\ff\phi(\3n+\e_3,t)\ff\phi(-\3n,t) \nonumber \\
  O_{\e_3,3}(t) &=& \ff\phi(2\e_3,t)\ff\phi(-\e_3,t) \la{operator-1} \\
  O_{\e_3,4}(t) &=& \fr14\sum_{\3n} \delta_{n_3,0}\,\delta_{n^2,2}\,
		\ff\phi(\3n+\e_3,t)\ff\phi(-\3n,t) \nonumber \,.
\ea
The ordering of the operator is such that the free particle energy
levels increase with increasing operator numbers.  The summation over
$\3n$ in the above operators makes them cubically ($d=0$) and
tetragonally ($d=1$) invariant.  This enhances their coupling to the
states which belong to the $A_1^+$ representation of the cubic and
tetragonal groups (see section~\ref{sec:energy} for details).

Let us now define the $\rho$ field operator
$O_{\3d,0}(t)=\ff\rho(\3d,t)$, where $|\3d|=0$ or 1.  With these
operators we measure the correlation function matrices
\ba
  C_{\alpha\beta}^0 (t) &=& \left\lx
	O_{0,\alpha}(t)\, O_{0,\beta}(0) \right\rx_c
  \la{c-0con} \\
  C_{\alpha\beta}^1 (t) &=& \fr13 \sum_{i=1,2,3}
	\left\lx O^*_{\e_i,\alpha}(t) \, O_{\e_i,\beta}(0)\right\rx_c
  \la{c-1}
\ea
where $\alpha,\beta = 0,1,2,\ldots$, and $\lx\cdot\rx_c$ denotes the
connected part of the correlation function.  Both of the matrices are
real and symmetric; $C^0_{\alpha\beta}$ by construction and
$C^1_{\alpha\beta}$ by statistical averaging (which can be made
explicit with the relation $O_{\e,\alpha} = O^*_{-\e,\alpha}\,$).

The correlation functions entering the matrix $C^1_{\alpha\beta}$ have
vanishing disconnected contributions.  The disconnected part of the
correlations in the matrix $C_{\alpha\beta}^0$ can be excluded by
considering instead of the function \nr{c-0con} the correlation
function \cite{Luscher90}
\be
  C^0_{\alpha\beta} (t) = \left\lx[O_{0,\alpha}(t) -
	O_{0,\alpha}(t+1)]\, O_{0,\beta}(0)\right\rx.
\la{c-0}
\ee
The spectral decomposition of the matrices $C^0$ and $C^1$ has the
form
\be
  C^d_{\alpha\beta} = \sum_{a} v_\alpha^{d,a*} v_\beta^{d,a}
	e^{-tW^d_a} \h\h v_\alpha^{d,a} = c^d_a \lx a|
	O^d_\alpha(0)|0\rx \,,
\ee
where $a$ labels the energy eigenstates (of energy $W^d_a$), $d$ is
either $0$ or $1$, and the coefficient $c^0_a = (1-e^{-W^0_a})^{1/2}$
and $c^1_a = 1$.

The energies $W^d_a$ can now be solved from the eigenvalue equation
$C^d(t) \psi^d_a = \lambda^d_a(t)\psi^d_a$, where we expect
$\lambda^d_a(t)\sim e^{-tW^d_a}$ when $t$ is large enough.  However,
the signal becomes exponentially smaller when $t$ increases, and is
soon lost into statistical noise.  The better the set of operators
used in the $C$-matrices spans the space of the lowest energy
eigenstates, the faster the convergence is.  The number of operators
should be larger than the number of energy eigenstates in the elastic
regime $2\m\le W<4\m$.  A larger set provides a better representation
of the eigenstates at the expense of increased numerical noise.

L\"uscher and Wolff \cite{Luscher90} proposed a method that allows a
reliable determination of the energy levels even at small values of
$t$.  Let us define the generalized eigenvalue problem (we drop the
label $d$ in the following formulae; they are valid for both $d=0$ and
$d=1$)
\be
  C(t) \psi_a = \lambda_a(t,t_0)\,C(t_0) \psi_a
\ee
where $t_0$ is a small reference time ($t_0=0$, for example).  The
eigenvalues are given by
\be
  \lambda_a(t,t_0) = e^{-(t-t_0)W_a}
  \la{expfit}
\ee
up to negligible corrections exponential in $t-t_0$.  Note that
$\lambda_a(t_0,t_0) = 1$ by construction.  This method can be
implemented by studying the (standard) eigenvalue problem of matrices
\be
  D(t) = C^{-1/2}(t_0) C(t) C^{-1/2}(t_0)
\la{d-matrix}
\ee
which has eigenvalues $\lambda_a(t,t_0)$ and eigenvectors $\Psi_a =
C^{1/2}(t_0)\psi_a$.

In our simulations we always measure the $6\times6$ ($d=0$) and
$5\times5$ ($d=1$) matrices as defined in eqs.~\nr{c-0} and
\nr{c-1}.  In the case of smallest volumes ($L\le 10$--12) we drop
the highest momentum correlation functions before analyzing the
eigenvalues.  For the largest volumes, there are more states below
elastic threshold than the dimension of the matrices; nevertheless, we
could always find the lowest 3--4 eigenstates without problems.

\subsection{Energy spectrum}

We measure the correlation function matrices $C^0(t)_{\alpha\beta}$
and $C^1(t)_{\alpha\beta}$, \eqs\nr{c-0} and \nr{c-1}, up to the
distance $T/2$; all the measurements are blocked into 35 -- 65 blocks,
which are used as independent measurements for jackknife error
analysis.  To obtain the transformed correlation matrix $D(t) =
C^{-1/2}(t_0)C(t)C^{-1/2}(t_0)$ we choose the reference time $t_0=0$;
in our checks the results were virtually unchanged for other (small)
choices of $t_0$.

We diagonalize the matrices $D(t)$ and order the eigenvalues
$\lambda_a(t)$, $t>0$, so that $\lambda_1 >\lambda_2 >\ldots
>\lambda_n$.  The energy spectrum is determined by a fully correlated
one-parameter exponential fit $\lambda_a(t) = \exp(-W_at)$.  The fit
range varies from $t=0$--10 to 0--4; the end point of the range is
adjusted until an acceptable $\chi^2$ per degree of freedom is
obtained.  Generally, the length of the range can be chosen longer
with lower energy states and larger lattices.  The effect of using fit
functions of correct periodicity has a negligible effect on the energy
spectrum (for correlations $D^0(t)$ the correct form is
$\propto\exp[-W_at] -\exp[-W_a(L-t-1)]$, due to the definition
\eq\nr{c-0}; for correlations $D^1(t)$ it is $\propto\exp[-W_at] +
\exp[-W_a(L-t)]$).

When the total momentum $\3P = 2\pi\e_i/L$ the eigenvalue analysis
yields only the laboratory frame energies $W_{\lf,a}$.  These have to
be transformed into the center of mass frame.  In the continuum, the
relation between the center of mass and laboratory frame energies is
given by
\be
  W_{\cm,a} = \sqrt{W_{\lf,a}^2 - P^2}
\la{lf-to-cm}
\ee
(see \eq\nr{wlevels}).  However, this relation is affected by the
finite lattice spacing, and motivated by the lattice dispersion
relation \nr{dispersion} we define
\be
  \cosh W_{\cm,a} = \cosh W_{\lf,a} - (1 - \cos P).
\la{defwcm}
\ee
The use of this relation is also justified by the excellent agreement
between the results of $P=0$ and $P=2\pi/L$ when the phase shift
function is determined (section \ref{sec:phaseresult}); if the
relation \nr{lf-to-cm} is used instead, there remains a clear
systematic difference between the momentum sectors.

\begin{figure}[tb]
\vspace{-1.2cm}
\epsfxsize=11cm
\centerline{\hspace{-1cm}\epsfbox{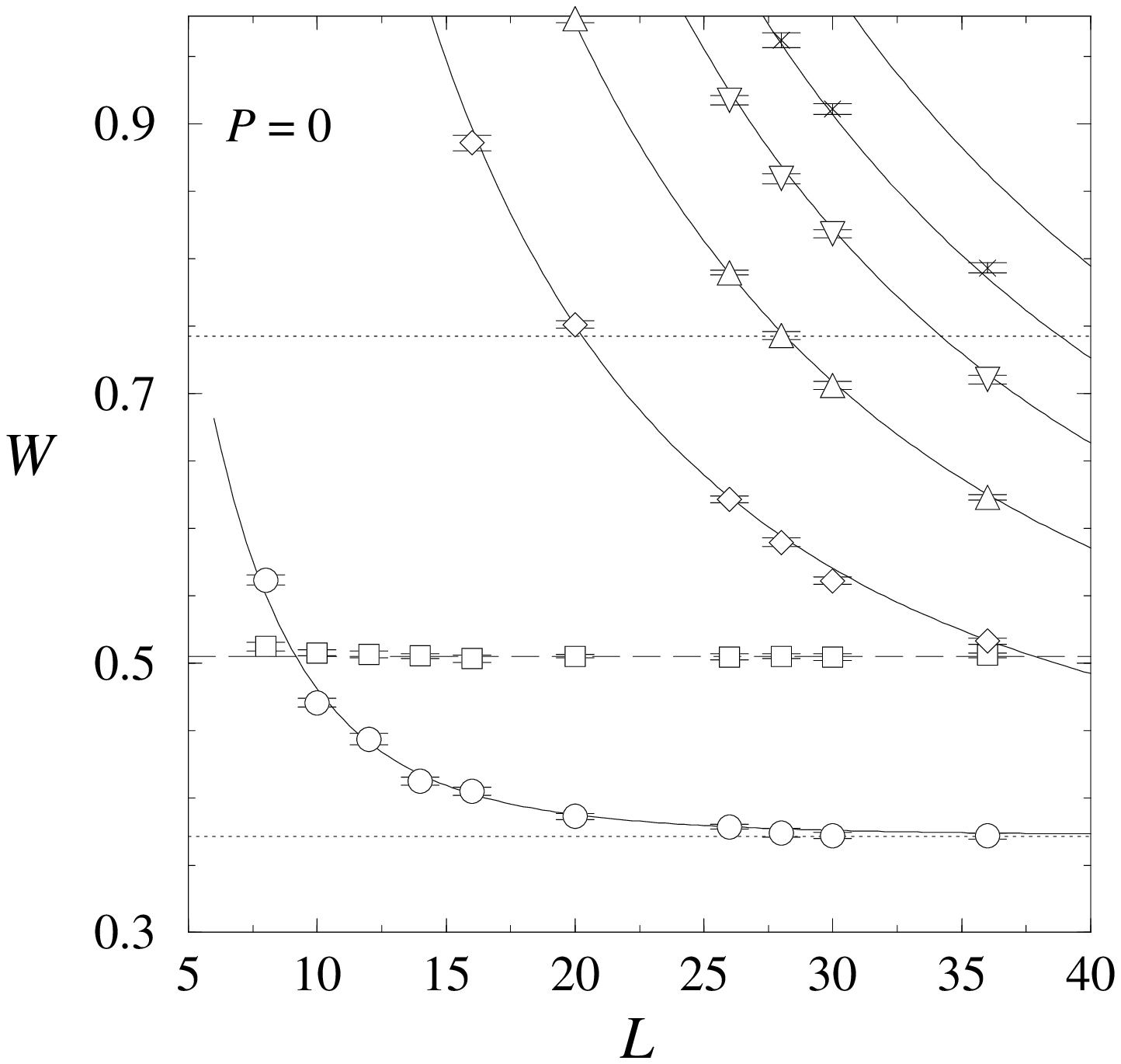}\hspace{-3.8cm}
\epsfxsize=11cm\epsfbox{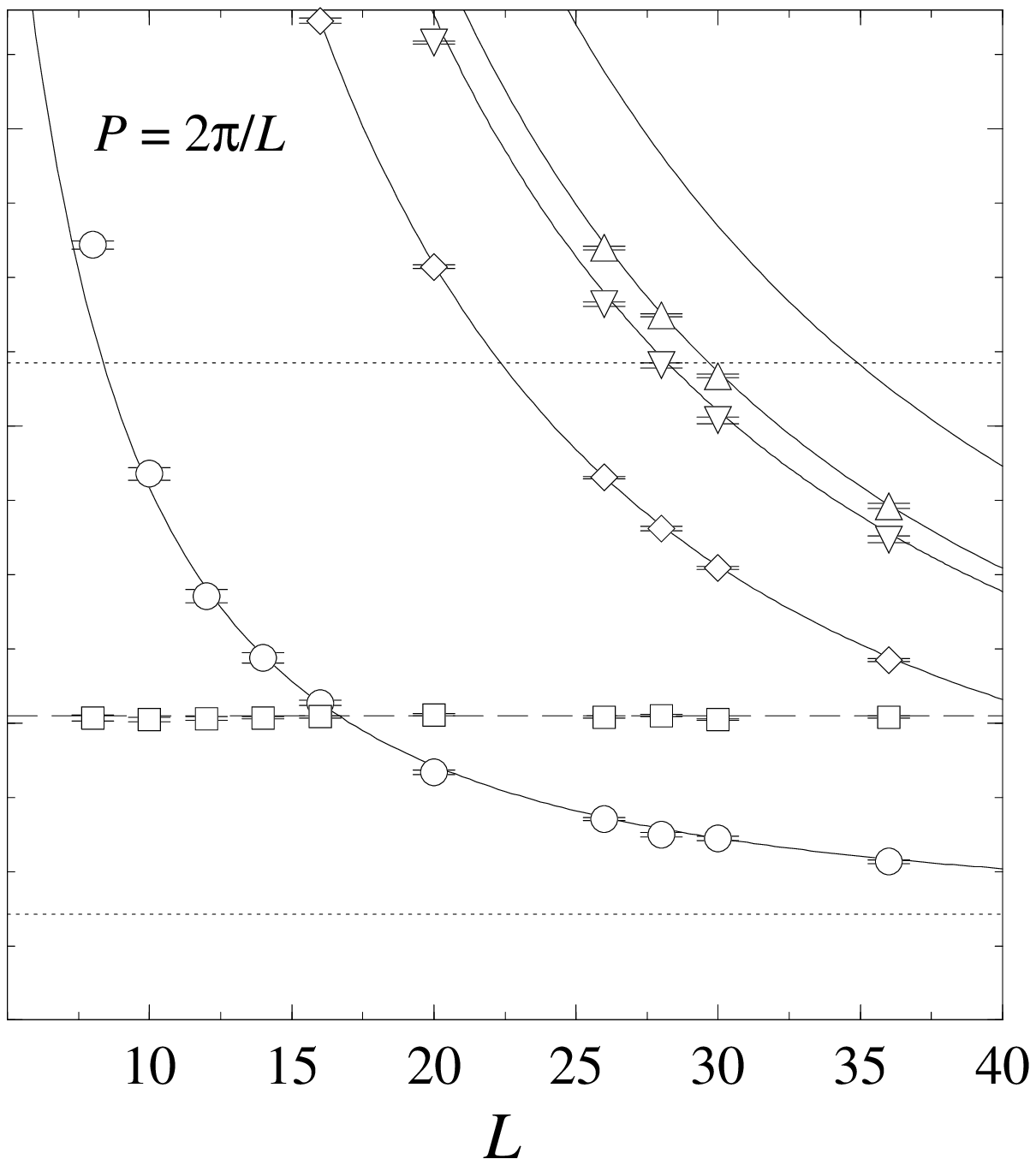}}
\vspace{-5.8cm}
\caption[0]{The center of mass frame energy levels for case A
($\kx=0$) for $P=0$ and $P=2\pi/L$.  The horizontal dotted lines show
the energy levels $2\m$ and $4\m$, and the dashed line corresponds to
$W=\M$.  The continuous curves are perturbative energy levels fitted
to the phase shift function
(sec.~\ref{sec:phaseresult}).\la{fig:wlevelsA}}
\end{figure}

\begin{figure}[tb]
\vspace{-1.2cm}
\epsfxsize=11cm
\centerline{\hspace{-1cm}\epsfbox{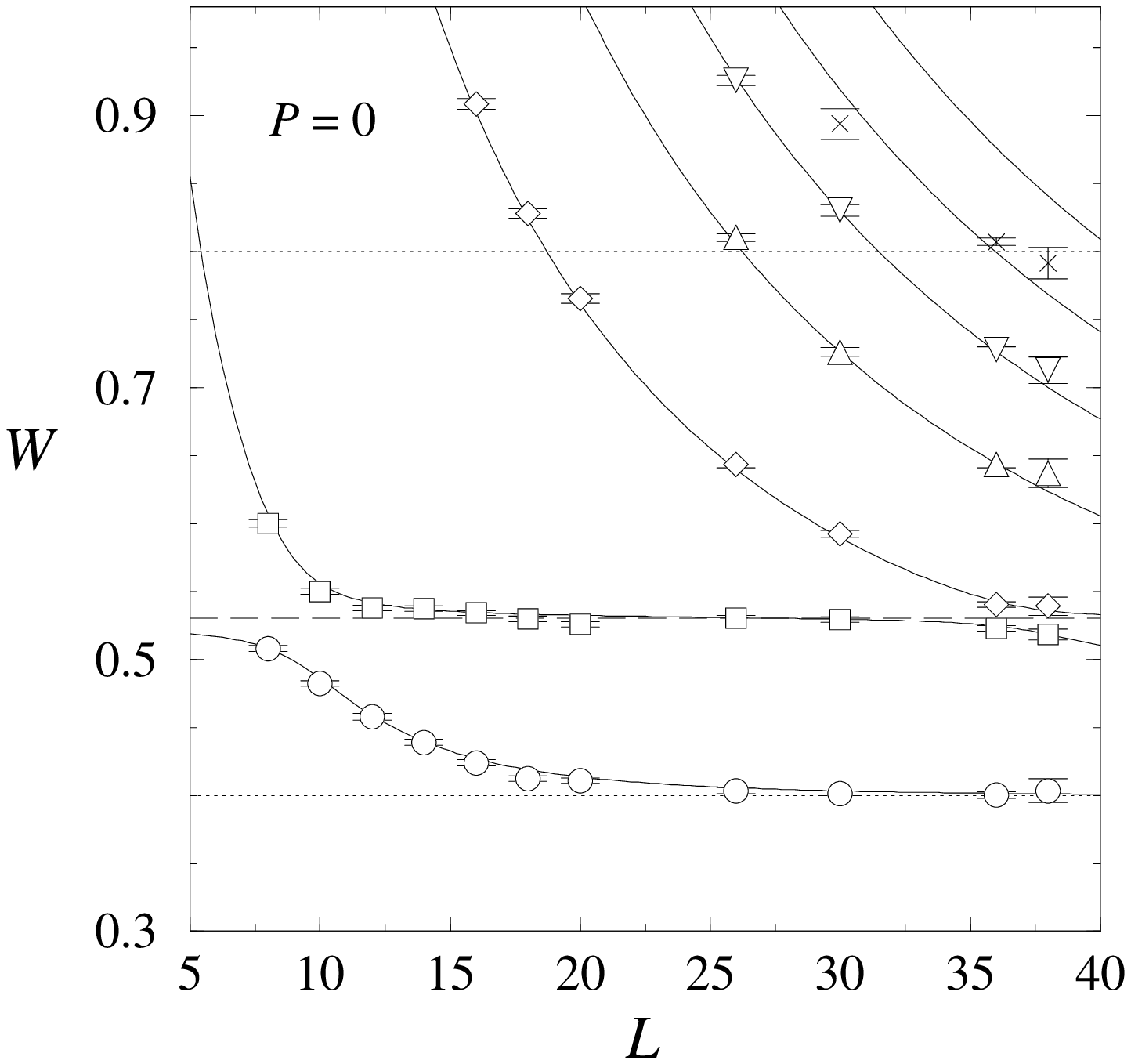}\hspace{-3.8cm}
\epsfxsize=11cm\epsfbox{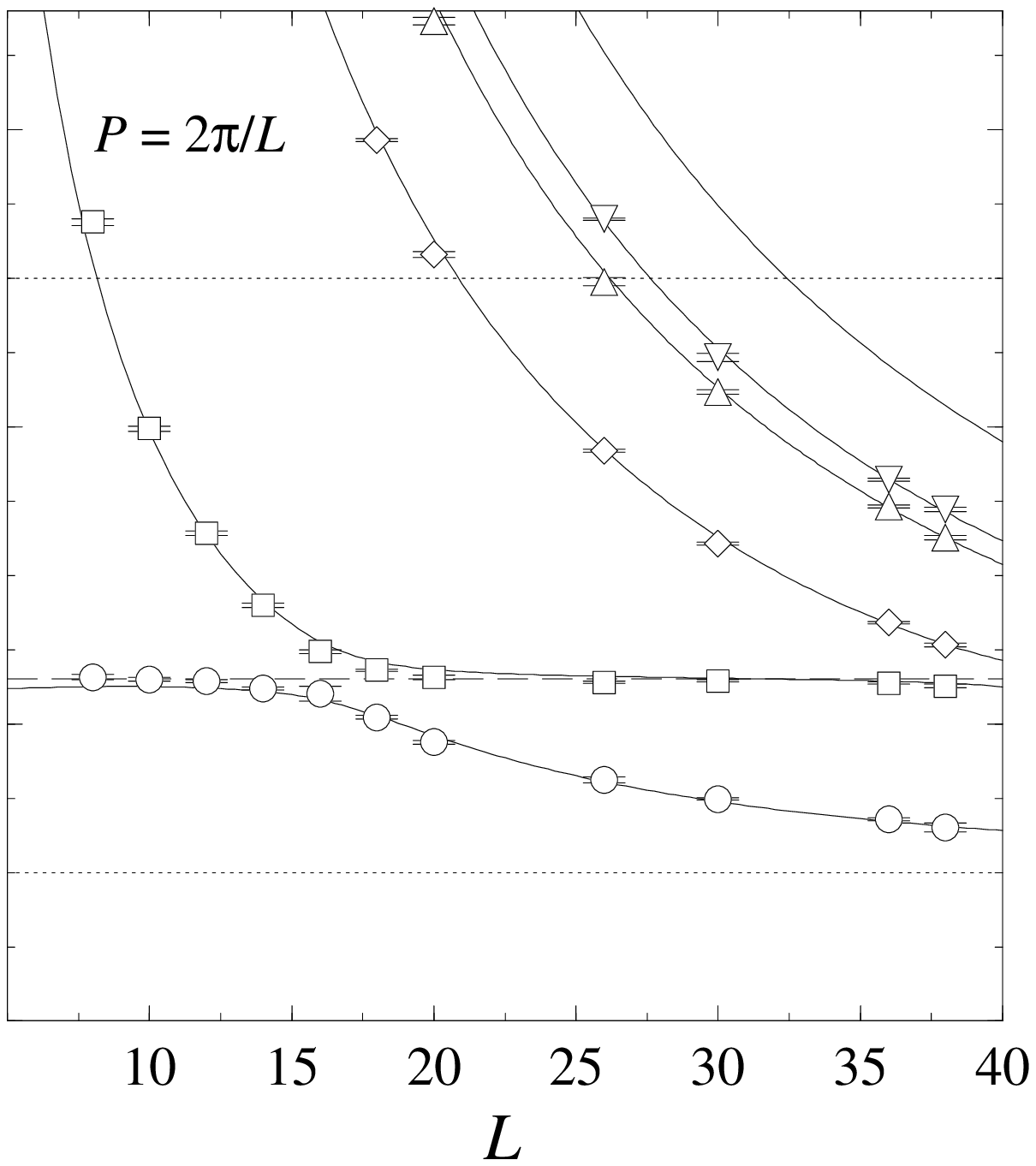}}
\vspace{-5.8cm}
\caption[0]{The center of mass frame energy levels for case B
($\kx=0.008$).  The resonance mass $\M$ is shown with a dashed line,
levels $W=2\m$ and $4\m$ with dotted lines, and the perturbative fit
with continuous lines.\la{fig:wlevelsB}}
\vspace{-0cm}
\end{figure}
\begin{figure}[tb]
\vspace{-1.2cm}
\epsfxsize=11cm
\centerline{\hspace{-1cm}\epsfbox{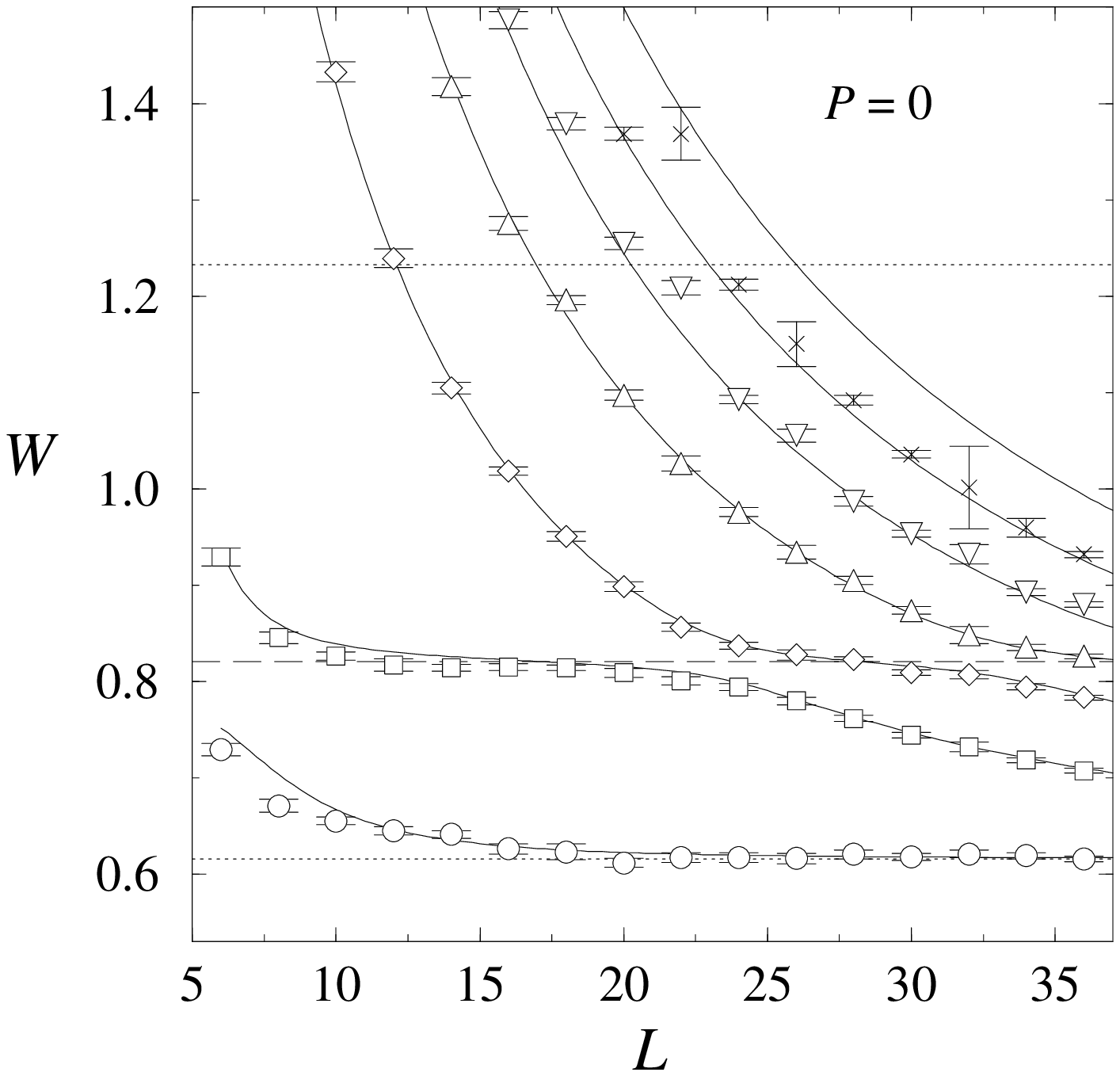}\hspace{-3.8cm}
\epsfxsize=11cm\epsfbox{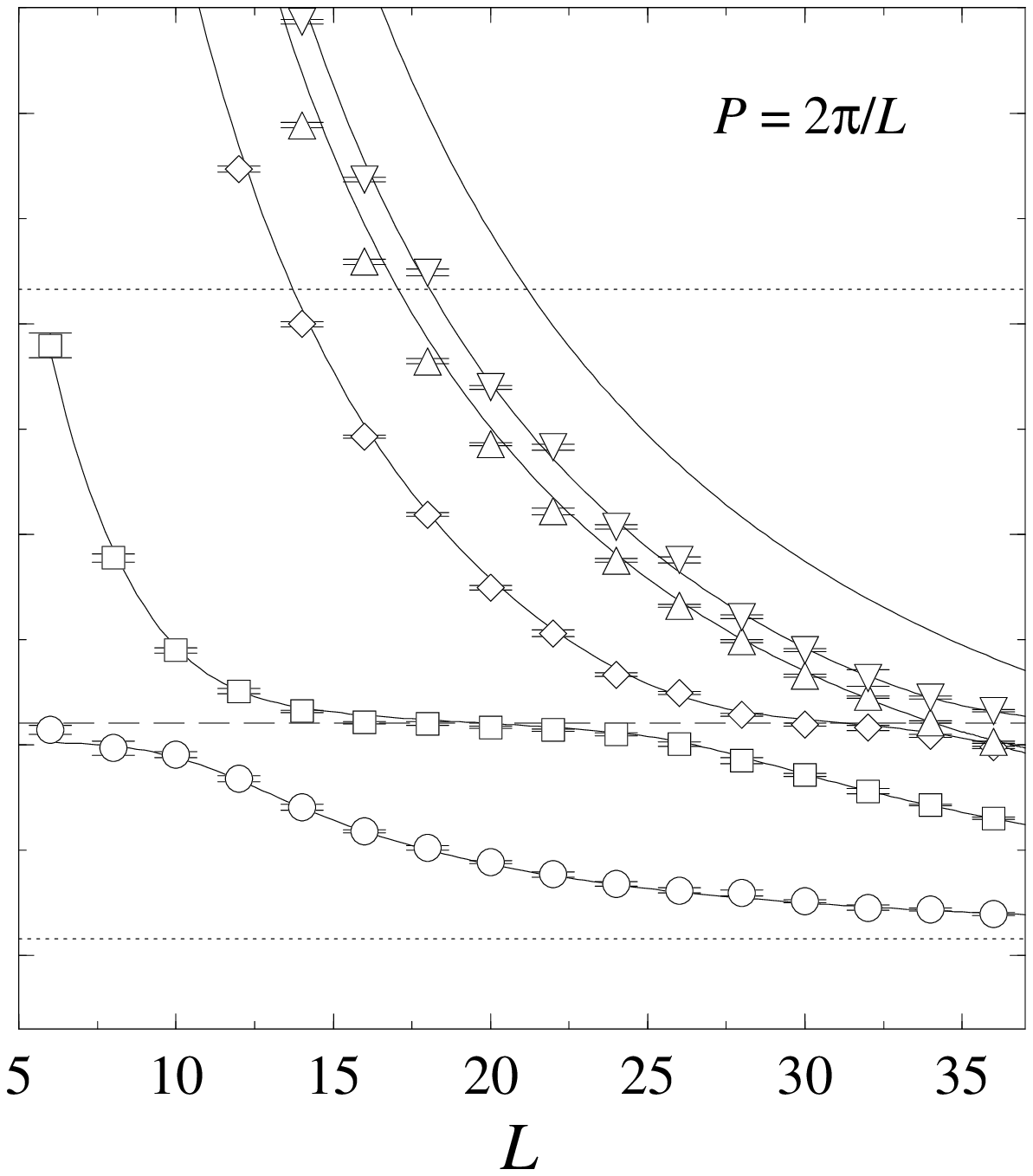}}
\vspace{-5.8cm}
\caption[0]{The center of mass frame energy levels for case C
($\kx=0.021$).\la{fig:wlevelsC}}
\vspace{-0.3cm}
\end{figure}

The center of mass energy levels are shown in figures
\ref{fig:wlevelsA}, \ref{fig:wlevelsB} and \ref{fig:wlevelsC}
for cases A, B and C, respectively, and for both momentum sectors
$P=0$ and $P=2\pi/L$ (left and right figures).  The energy level
$W=2\m$ and the inelastic threshold $W=4\m$ are shown with horizontal
dotted lines.\footnote{In the $P=2\pi/L$ case, the inelastic limit is
actually {\em greater\,} than $4\m$: when we ignore the interactions,
in the laboratory frame the threshold is $W_{\lf,\rm inel.} = 3\m +
[\m^2 + (2\pi/L)^2]^{1/2}$, corresponding to 3 particles at rest and
one moving with momentum $p=2\pi/L$.  In the cm-frame this yields
$W^2_{\cm,\rm inel.} = 10\m^2 + 6\m[\m^2 + (2\pi/L)^2]^{1/2} >
(4\m)^2$.}  Above the inelastic limit new 4 particle states appear and
our analysis loses its validity (it should be noted that the 3
particle states do not couple to the 2 particle ones due to the lack
of the $\phi^3$ coupling).

The solid lines in figures \ref{fig:wlevelsA}--\ref{fig:wlevelsC} are
perturbative energy levels determined by fitting equations
(\ref{deltasum}--\ref{singular}) to the phase shift data.  In all
cases the consistency of the fits is excellent, although the $\chi^2$
values are not necessarily very good.  The extraction of the phase
shift data and the fitting procedure are described in section
\ref{sec:phaseresult}.

In case A the $\phi^2\rho$ coupling is absent and the $\rho$ particle
is stable.  The energy level $W=\M$ is shown with a horizontal dashed
line in fig.~\ref{fig:wlevelsA}.  In cases B and C $\M$ is determined
by the fit to the perturbative ansatz \eq\nr{singular} and shown in
figs. \ref{fig:wlevelsB} and \ref{fig:wlevelsC} with dashed lines.

Qualitatively, the essential difference between the non-resonance case
(A) and the resonating cases (B and C) is obvious: the intersections
of the $2\phi$ levels with the $\rho$ energy level in case A turn into
avoided level crossings in cases B and C.  It is especially
illuminating to consider cases A and B, which correspond to nearly
identical values of $\m$ and $\M$: when the total momentum $P=0$, the
lowest level crossing occurs at $L\sim 9$.  This level crossing,
absent in the free theory (see fig.~\ref{fig:energylevels}), is caused
by the repulsive $\lambda\phi^4$ interaction which increases the
energy of the lowest 2 particle state.  If $\lambda< 0$, the energy of
the lowest state decreases with decreasing lattice size, and the first
level crossing vanishes (This effectively occurs in O(4)-symmetric
$\phi^4$ theory in external field \cite{Gockeler94}).

In the $P=2\pi/L$ sector the first (avoided) level crossing has moved
up to $L\sim 16$.  This crossing is determined by the kinematics of
the $\phi$ and $\rho$ particles, and it persists irrespective of the
value of $\lambda$.  In our case this is the first ``useful'' level
crossing, since the first crossing in the $P=0$ sector occurs at such
a small volume ($L<2\m^{-1}$) that the reliability is compromised by
the finite volume polarization effects (nevertheless, to our surprise
the data and the perturbative fits agree quite well even down to the
smallest volumes --- which were not used to determine the fit).

In case C the parameters were chosen so that the values of $\m$ and
$\M$ are $\sim 50\%$ larger than in A and B\@.  This allows us to
observe 3 avoided level crossings in the range $L=6$--36 as opposed to
2 in case B\@.  The coupling constant $\kx$ is also considerably
larger (0.021 vs. 0.008), giving a broader resonance.

We also observe the appearance of the closely spaced energy levels
(numbers 4 and 5 from the lowest up) in the $P=2\pi/L$ case.  In the
free theory these correspond to the laboratory frame momentum states
$\3q_1=(0,0,2)$, $\3q_2=(0,0,-1)$ and $\3q_1=(1,1,1)$,
$\3q_2=(0,-1,-1)$, respectively (here $\3q_i=\3p_i\,2\pi/L$); the
degeneracy of these states is lifted by relativistic effects (see the
discussion in section \ref{sec:noninteract}).

\begin{figure}[tb]
\vspace{-1.2cm}
\epsfxsize=11cm
\centerline{\epsfbox{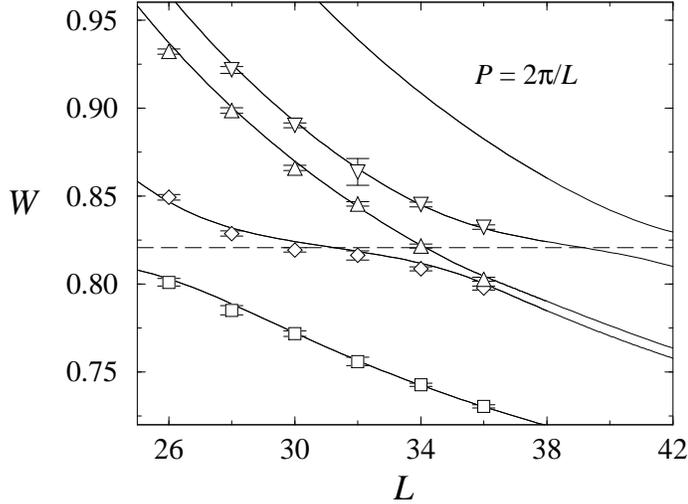}}
\vspace{-7.1cm}
\caption[0]{Close-up of the $P=2\pi/L$ sector energy levels in
case C\@.  The energy levels shown are numbers 2 (box), 3 (diamond), 4
(triangle up) and 5 (triangle down), counted from the lowest energy
state up.\la{fig:wlevels-fine}}
\end{figure}

In case C the largest volume is just large enough so that the
$P=2\pi/L$ levels 4 and 5 are in the middle of the avoided level
crossing region.  A close-up view of this region is shown in figure
\ref{fig:wlevels-fine}.  At volumes smaller than $\sim 34^3$ the
levels 4 and 5 are close to each other, and at larger volumes the
levels 3 and 4 are nearly degenerate.  Even in this complicated region
the agreement between the data and the perturbative fit is remarkable,
especially since none of the data of the levels 3, 4 and 5 is used to
determine the fit parameters.

Generally, the statistical errors in the $P=2\pi/L$ sector tend to be
smaller than in the $P=0$ sector.  This is probably due to two
reasons: the correlation functions in the $P=2\pi/L$ sector do not
have a disconnected part which needs to be subtracted (cf. equations
\nr{c-0} and \nr{c-1}), and the statistics is increased by combining the
measurements in $\3P=2\pi\e_i/L$, $i=1,2,3$ sectors together.  While
not being completely statistically independent --- the operators with
different directions of $\3P$ still act on the same configurations ---
the operators belonging to different sectors are orthogonal;
\ie the cross-correlations between operators belonging in
different $\3P=2\pi\e_i/L$ sectors vanish.

\subsection{Scattering phase shift\la{sec:phaseresult}}

%\subsubsection{Lattice effects}
In order to calculate the phase shift $\delta_0(p)$ from the
fundamental equations (\ref{deltal}--\ref{zetaf00}), we need to
transform the energy eigenvalues $W_a$ to the corresponding momenta
$p_a$.  In continuum, the relation is given by \eq\nr{wcm}: $p_a^2 =
(W_a/2)^2 - \m^2$.  However, as was observed in the analysis of the
single particle energy spectrum in section \ref{sec:single}, the
lattice dispersion relation
\be
  \cosh\fr{W_a}{2} - \cosh\m = \sum_{i=1}^3 (1-\cos p_{a,i})
\la{dispW}
\ee
can be expected to give a much more accurate relation when the lattice
spacing is finite.  However, due to the breaking of the translational
and rotational symmetries on the lattice the assignment
$W_a\rightarrow p_a(W_a)$ is not uniquely defined by \eq\nr{dispW}.
This problem can be circumvented by various approximations: since we
are interested in the elastic region $0\le p_a<\sqrt{3}\m$, the
magnitude of $p_a$ is rather small (in lattice units) and it is
bracketed by
\be
  \sqrt{3}\cos^{-1}(1-\fr13 x) \le p_a \le \cos^{-1}(1-x)
\la{limits}
\ee
where $x\equiv\cosh(W_a/2)-\cosh\m$.  The first limit is saturated
when $p_1=p_2=p_3=p/\sqrt{3}$, and the second when $p_1=p$,
$p_2=p_3=0$.  The difference is maximal at the inelastic limit $W_a =
4\m$, and using $\m=0.3$ ($\sim$ our largest value, case C) the limits
become $0.5315\le p_a(4\m)\le 0.5358$: the uncertainty in $p_a$ is
always smaller than one percent, and in the important region
$W_a\sim\M$ it is still smaller.  Either of the limits above can be
used to calculate $p_a(W_a)$; however, the statistical quality of the
data is good enough that the results are not compatible (\ie not
within one standard deviation of each other).

We refine the estimate of $p_a$ as follows: we find two consecutive
{\em free particle} energy levels $W_n$ and $W_{n+1}$ so that $W_n\le
W_a\le W_{n+1}$.  These energy levels are calculated from the lattice
dispersion relation \nr{dispW} using lattice momenta $\3p =
(2\pi/L)\3m$ in the $P=0$ sector and $\3p = (2\pi/L)\,\g(\3m +
\e_3/2)$ in the $\3P=2\pi\e_3/L$ sector\footnote{The value of $\gamma$
in the $P=2\pi/L$ sector is calculated from $\gamma = W_\lf/W_\cm$;
see \eqs(\ref{defgamma}--\ref{wlfwcm}).} ($\3m\in\Z^3$).  The desired
momentum vector $\3p_a$ is then found by interpolating $\3p_a = r\3p_n
+ (1-r)\3p_{n+1}$, $0\le r\le 1$, until \eq\nr{dispW} is satisfied.
This method presupposes that the eigenstates with energies in between
two free levels predominantly consist of a mixture of the two free
states.  This is supported by the analysis of the eigenvectors at the
end of this section.

\begin{figure}[tb]
\vspace{-1.2cm}
\epsfxsize=11cm
\centerline{\epsfbox{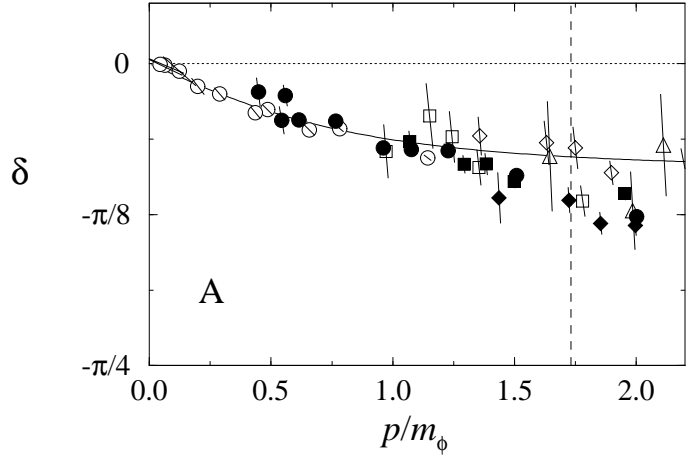}}
\vspace{-8cm}
\caption[0]{The phase shift for case A ($\kx=0$).  Open and filled
symbols belong to sectors $P=0$ and $P=2\pi/L$, respectively.  The
inelastic limit $p=\sqrt{3}\m$ is shown with a vertical dashed line,
and the solid line is a perturbative fit.\la{fig:phaseA}}
\end{figure}

The phase shift is calculated by inserting the momenta $p_a$ to
\eqs(\ref{deltal}--\ref{zetaf00}).  In figure \ref{fig:phaseA}
we show the phase shift of case A as a function of the center of mass
momentum for both $P=0$ and $P=2\pi/L$ sectors with open and filled
symbols, respectively.  The tilted lines are one standard deviation
errors; since $p_a$ completely determines $\delta_0$ by \eq\nr{deltal}
the errors in the values of $p_a$ and $\delta_0$ are fully correlated.

\begin{figure}[tb]
\vspace{-1.2cm}
\epsfxsize=11cm
\centerline{\hspace{-1cm}\epsfbox{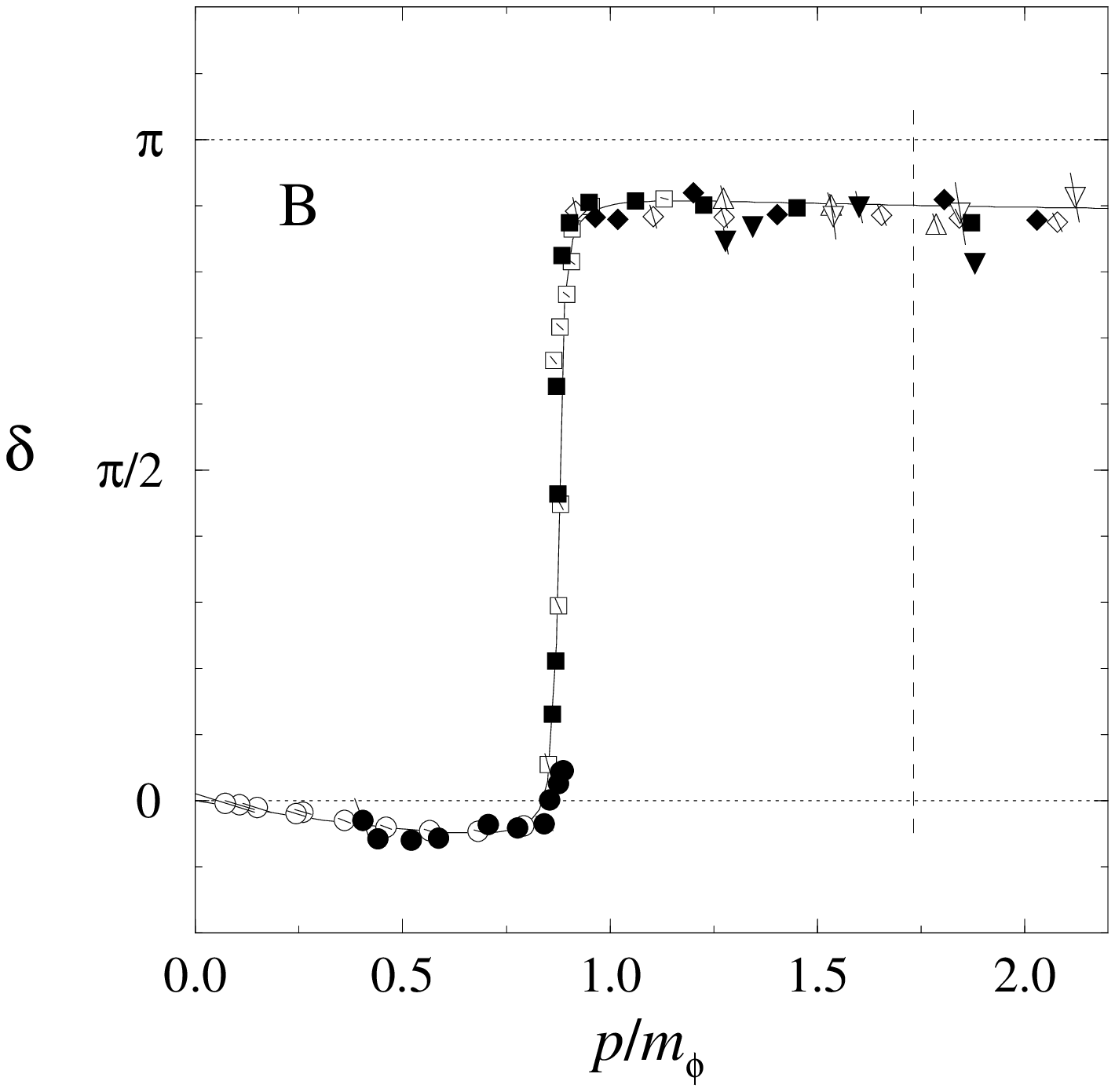}\hspace{-3.8cm}
\epsfxsize=11cm\epsfbox{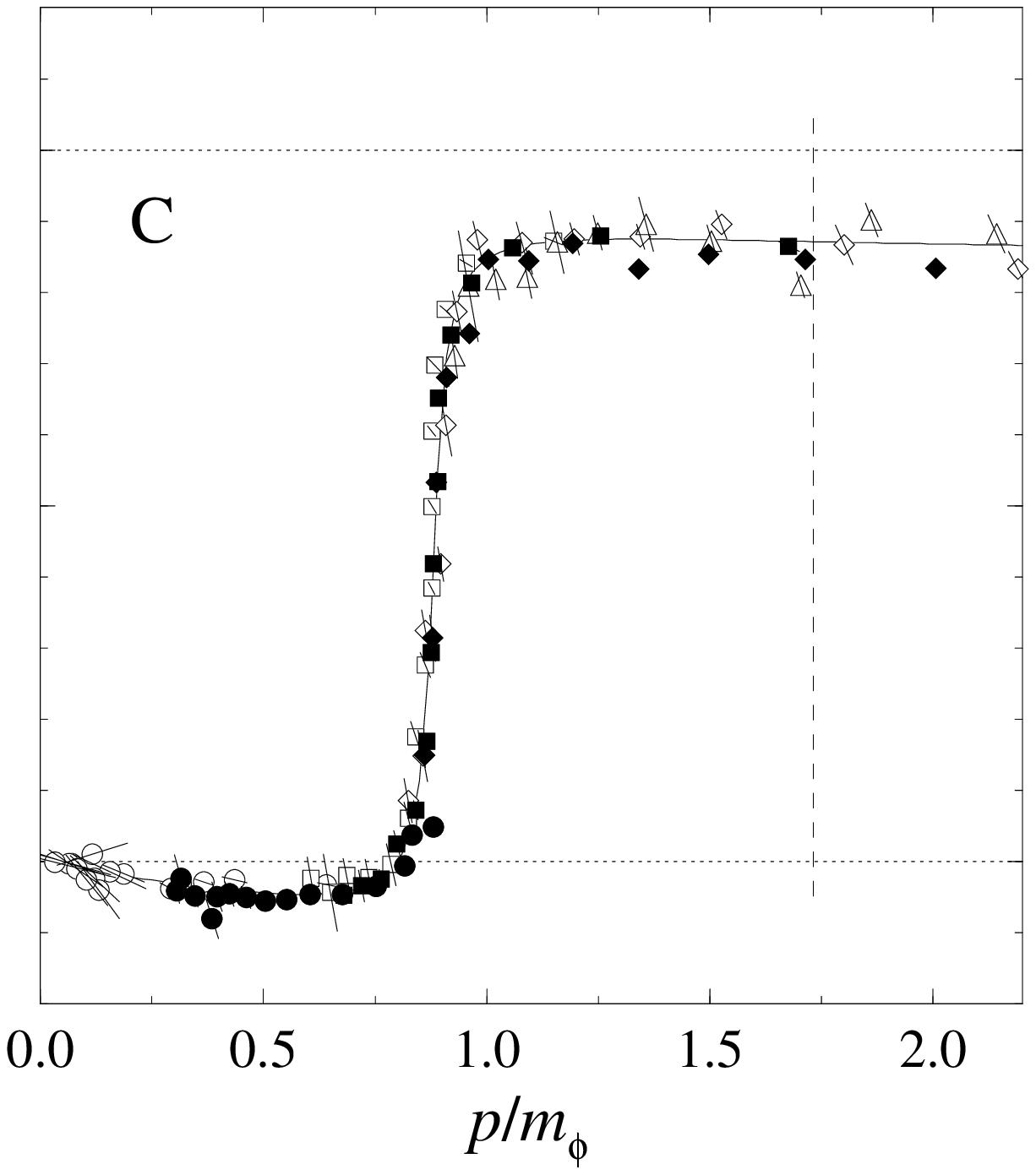}}
\vspace{-5.8cm}
\caption[0]{The phase shift for cases B ($\kx=0.008$) and C
($\kx=0.021$).\la{fig:phaseBC}}
\end{figure}

The phase shifts of cases B and C are shown in figure
\ref{fig:phaseBC}.  The emergence of the resonance is evidenced by
the rapid increase in $\delta_0$ by $\pi$ when
$p^2+\m^2\approx(\M/2)^2$ (Note the difference in the vertical scales
in figures \ref{fig:phaseA} and \ref{fig:phaseBC}).  In figures
\ref{fig:phaseA} and \ref{fig:phaseBC} the symbol shapes
are the same as in the energy level figures
\ref{fig:wlevelsA}--\ref{fig:wlevelsC}.

The consistency between the two total momentum sectors is good in all
three cases.  However, one can observe some systematic differences
between the phase shift calculated from the data from different energy
levels, especially at large momenta ($p\sim p_{\rm inelastic}$).  This
is likely due to some unaccounted for lattice effects.

\subsubsection{Resonance parameters}

The tree level perturbative calculation gives the scattering phase
shift in the $l=0$ sector as $\delta_0(p) = \delta_r(p)+\delta_s(p)$,
where
\ba
\delta_r &=& -\lambda_\1R \fr{p}{16\pi W} +
	\fr{\kx_\1R^2}{32\pi}\fr{1}{Wp}
	\log\fr{4p^2+\M^2}{\M^2} \la{dreg} \\
\tan\delta_s &=&  -\fr{\kx_\1R^2}{16\pi}\fr{p}{W}
 	\fr{1}{W^2-\M^2} \,. \la{dsing}
\ea
Equations \nr{dreg} and \nr{dsing} are fitted to the phase shift data.
In case A, $\kx_\1R=0$ and there is only one fit parameter
$\lambda_\1R$, in cases B and C we fit the parameters $\lambda_\1R$,
$\kx_\1R$ and $\M$.  In order to reduce finite volume polarization and
finite lattice spacing effects, we exclude points with $p >1.5\m$ and
small lattice volumes.  Only two lowest energy levels from each $P$
sector are used, except in case C where we use 3 lowest levels from
the $P=0$ sector.  Both of the momentum sectors are fitted together.
We take it as a sign of the consistency of the calculation that the
results do not vary significantly if only one of the sectors is
included in the fit.  The $\chi^2$ values of the fits are 21/19
d.o.f.\@ for case A, 72/32 d.o.f.\@ for B and 63/60 d.o.f.\@ for C\@.
The large $\chi^2$ value for case B is partly due to the very sharp
nature of the resonance: $\delta_l(p)$ varies very rapidly in a narrow
$p$-range, and even small unaccounted systematic factors can have a
large effect.

The fitted parameters are listed in table \ref{table:runs}, and the
fitted functions are shown in figures \ref{fig:phaseA} and
\ref{fig:phaseBC} with solid lines.  Case A corresponds to pure
lattice $\phi^4$ theory, and the fitted value for $\lambda_\1R$ is
$28.1\pm 1.1$.  This is slightly larger than the perturbative value
$\lambda_\1R=24\pm 3$ at the bare lattice coupling $\kp=0.0742$
\cite{Luscher87}.  However, if we include only the points with
$p<0.6\m$ in the fit, the result is $\lambda_\1R=26\pm2$.

In cases B and C $\lambda_\1R$ is $36.8\pm 1.3$ and $48.3\pm 2.0$,
respectively.  These are close to but still somewhat larger than the
perturbatively calculated pure lattice $\phi^4$ theory values $29\pm
3$ at $\kp=0.07325$ and $42\pm 7$ at $\kp=0.07075$ \cite{Luscher87}.
In this case one can expect the values to be different, since the
coupling to the $\rho$ field affects the renormalization.

The value of the renormalized 3-point coupling $\kx_\1R$ is
$0.598(14)$ (B) and $1.49(3)$ (C).  The ratio of these values is
$0.40(1)$, which is quite close to the ratio of the bare couplings
$0.008/0.021\approx 0.38$.

The resonance masses are $\M=0.5306(13)$ in case B and $0.8206(11)$ in
case C\@.  The resonance widths $\G$ are calculated from
\be
  \G = \fr{\kx_\1R^2}{32\pi\M^2}\sqrt{\M^2 - 4\m^2}\,,
\ee
with the results $\G=0.0044(2)$ (B) and 0.0178(7) (C).  In both cases
the resonance is quite narrow: in case B, the ratio $\G/\m$ is only
0.8\% and in case C 2.2\%.

An alternate method to calculate the resonance mass and width is to
use the generic Breit-Wigner relativistic form
\be
  \tan\delta(p) = \fr{\M\G}{\M^2 - W^2}\,.
  \la{Breit-Wigner}
\ee
This is valid only in the small neighborhood of the resonance value
$\delta\sim\pi/2$.  For our data \eq\nr{Breit-Wigner} alone does not
give a good fit, but if we add the term proportional to $\lambda_\1R$
from \eq\nr{dreg} to the Breit-Wigner phase shift we obtain results
comparable to the perturbative ansatz (\ref{dreg}--\ref{dsing}).

\subsection{Wave function representation}

The decomposition of the energy eigenstates into states with definite
momentum tells us if the number of the states in the correlation
function matrices is large enough.  The projection to the momentum
states can be obtained from the vector $\psi_a = C^{-1/2}(t_0)\Psi_a$,
where $\Psi_a$ is an eigenvector of the matrix $D(t)$,
\eq\nr{d-matrix}.  After normalizing, the squares of the components
$(\psi_a^\alpha)^2$ give the relative contribution of momentum states
$\alpha$ to the energy eigenstate $a$.  The eigenvectors should be
independent of $t$, and typically they do not vary significantly until
the correlation function fitting range is exceeded.
\begin{figure}[tb]
\vspace{-1.2cm}
\epsfxsize=12.5cm
\centerline{\epsfbox{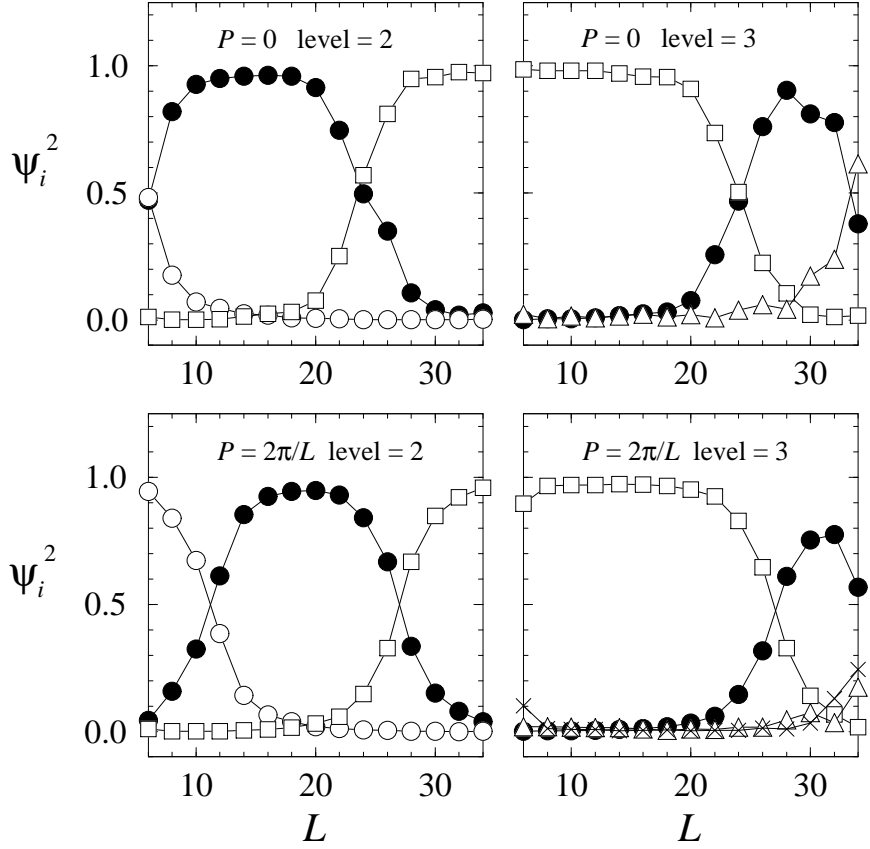}\hspace{1cm}}
\vspace{-4.3cm}
\caption[0]{The projections of the eigenstates to the $\rho$
and $2\phi$ momentum states in case C\@.  $\rho$ is shown with filled
circles, and $2\phi$ states with open circles, boxes, triangles and
crosses, from the lowest momentum value to the highest.  Only states
with an appreciable projection are shown.\la{fig:statesC}}
\end{figure}

In figure \ref{fig:statesC} we show the weights of the physical states
which belong to levels 2 and 3 in momentum sectors $P=0$ and
$P=2\pi/L$.  The values have been averaged over the range $1\le t\le
5$.  The contribution of $\rho$ is shown with filled circles.  In the
$P=0$ sector the $2\phi$ cm-momentum $p$ states are as follows: $q^2 =
(pL/(2\pi))^2 = 0$ is shown with open circles, $q^2=1$ with boxes, and
$q^2=3$ with triangles.  In the $P=2\pi/L$ sector we enumerate the
$2\phi$ states with $\3q_1$ and define $\3q_2 = (0,0,1)-\3q_1$.  The
levels shown are $\3q_1 = (0,0,1)$ with open circles, (0,1,1) with
boxes, (0,0,2) with triangles and (1,1,1) with crosses.  Only levels
with non-negligible amplitudes are shown.

The figure shows that away from the level crossing regions the
physical states are strongly dominated by only one momentum state.
Around the avoided level crossing the physical state is comprised of
mostly two momentum states with smoothly varying amplitudes.  An
exception to this rule occurs in the sector $P=2\pi/L$ when the level
3 approaches the avoided crossing with the nearly degenerate levels 4
and 5 ($L\sim 34$).  Both of the momentum states $\3q_1 = (0,0,2)$ and
(1,1,1) contribute to the physical state with non-negligible
amplitudes.

The strong dominance of only 1 or 2 momentum states in the
decompositions of the physical states indicates that the correlation
function matrices \nr{c-0} and \nr{c-1} have enough states to
accurately describe at least 3 lowest energy levels (which are the
only ones we used in the numerical analysis).  With the parameters
used in case C several new states start to approach the resonance mass
level after $L\sim 34$.  It becomes increasingly difficult to obtain
accurate information about the resonance: one needs correlation
function matrices with more states and with increased statistical
accuracy.  In case B this would happen at $L\sim 55$.

\section{Derivation of the phase shift formula\la{sec:phaseshift}}

In this section we derive the fundamental equation \nr{deltal}.  The
derivation uses the formalism derived in references
\cite{Luscher86,Luscher91}, generalizing it to encompass non-zero
total momentum case.  In order to keep the discussion simple, we are
working at the level of the relativistic 2-particle quantum mechanics.
The generalization to the field theory level can be done, for example,
with the Bethe-Salpeter formalism, as discussed by
L\"uscher~\cite{Luscher86}.

The expansion of the 2-particle scattering wave function in terms of
spherical harmonics has a physical meaning only in the center of mass
frame.  This is especially relevant in the case of resonance
scattering, where the resonance channel is an eigenstate of angular
momentum.  Also, we want to compare the results obtained with both
zero-momentum and non-zero momentum scattering sectors, and join the
data together in order to construct a more complete picture of the
scattering wave functions.  To reach this goal it is necessary for us
to be able to relate the laboratory frame measurements to the center
of mass quantities.

Due to the Bose symmetry the angular momentum $l=\mbox{odd}$ channels
have a vanishing scattering amplitude for identical spin-0 particles.
Therefore, in the following we shall implicitly assume that the fields
carry an extra quantum number making the particles distinguishable, if
necessary.  The projection to the desired symmetry sector can be made
at any stage of the analysis.

\subsection{Lorentz transformation of the wave function}

Let us consider two spinless bosons of mass $\m$ in an infinite
volume.  The state of the system is described by the scalar wave
function $\psi(x_1,x_2)$, where $x_i = (x^0_i,\3x_i)$ are the
4-dimensional Minkowski space-time coordinates of the particles.  The
wave function transforms in Lorentz transformations as
$\psi\rightarrow\psi'$:
\be
  \psi(x_1,x_2)=\psi'(x_1',x_2')=\psi'(\Lambda x_1,\Lambda x_2),
\la{psitransform}
\ee
where $(x')^\mu = {\Lambda^\mu}_\nu x^\nu$ denotes the Lorentz
transformation of the 4-vector $x$.  The wave function depends on two
time coordinates, and the space and time coordinates are mixed in
Lorentz transformations.  Often it is convenient to consider only the
case ${x_1}^0 = {x_2}^0$ (equal time formalism).  However, this
condition has different physical meanings in different inertial frames
and makes the formalism explicitly non-covariant.

The situation can be made simpler by using the special properties of
the center of mass frame of the particles.  Let us first consider {\em
non-interacting\,} particles: in any inertial frame, the wave function
satisfies the Klein-Gordon equations (we use here the metric tensor
sign convention $g_{\mu\nu} = {\rm diag}(1,-1,-1,-1)$)
\be
  ( \hat p_{i\mu} \hat {p_i}^\mu - \m^2 ) \psi(x_1,x_2) = 0,\h i=1,2
\la{KleinGordon}
\ee
where the standard operator relation $\hat p_{i\mu} =
-i\partial/\partial {x_i}^\mu$ is understood.  Equations
\nr{KleinGordon} can be transformed with the change of variables
\ba
  X &=& \half (x_1 + x_2)   \la{defX} \\
  x &=& x_1 - x_2           \la{defx}
\ea
into the form
\ba
  \left[4(\hat p_\mu\hat p^\mu - \m^2) +
	\hat P_\mu \hat P^\mu \right]
	\,\,\psi(x,X) &=& 0  \la{pequation}\\
  \hat p_\mu \hat P^\mu \,\, \psi(x,X) &=& 0    \la{pdotP}
\ea
where $\hat p = (\hat p_1 - \hat p_2)/2$, and $\hat P =\hat p_1 + \hat
p_2$ is the total 4-momentum operator.

In the absence of external potentials, the total momentum is conserved
and we can restrict ourselves to eigenfunctions of $P$ even when the
2-particle interaction is turned on:
\be
  \psi(x,X) = e^{-i P_\mu X^\mu} \phi(x).
\la{psiansatz}
\ee

The center of mass frame is defined as the frame where the space
components of the total momentum vanish: $\3P^*=0$ (the index * is
used to denote center of mass coordinate and momenta).  We shall
consider only positive kinetic energy solutions $P^*_0 = W_\cm > 2\m$.
{}From \eq\nr{pdotP} it follows that $\hat p^*_0 \phi_\cm(x^*) =
-i\partial_{0}\phi(x^*) = 0$.  Thus, in the center of mass frame the
wave function depends only on the time variable $t^* = X^{*0} =
\half(x_1^{*0} + x_2^{*0})$ and the relative separation of the particles
$\3x^*=\3x^*_1 -\3x^*_2$:
\be
  \psi_\cm(\3x^*,t^*) = e^{-i W_\cm t^*} \phi_\cm(\3x^*).
\la{psicm}
\ee

Let us now consider the situation in the laboratory frame.  The
transformation from the laboratory frame to the center of mass frame
can be written as $r^{*\mu} = {\Lambda^\mu}_\nu r^\nu$, where $r$ is
any position 4-vector and quantities without * refer to the laboratory
frame.  With the shorthand definition \eq\nr{defg}, this is
\ba
  r^{*0} &=& \gamma (r^0 + \3v\cdot\3r) \\
  \3r^*  &=& \g (\3r + \3v r^0)
\ea
where $\3v = \3P/P_0$ is the 3-velocity of the center of mass in the
laboratory frame.  Using the transformation
\nr{psitransform}, the identity $P_\mu X^\mu = P^*_\mu X^{*\mu}$ and
\eq\nr{psiansatz}, the laboratory frame wave function can be written
as $\psi_\lf(x,X)=\exp(-iP_\mu X^\mu) \phi_\lf(x)$, where
\be
  \phi_\lf(x) = \phi_\lf(x^0,\3x) = \phi_\cm(\g (\3x + \3v x^0))\,.
\la{philf}
\ee
Note that $\phi_\lf$ depends explicitly on $x^0 = {x_1}^0 - {x_2}^0$.
However, in the laboratory frame we are interested in the case where
both of the particles have equal time coordinate, $x^0 = 0$ (More
specifically, on the lattice all 2-particle operators act on single
spacelike hyperplane.).  In the \cm\ frame this corresponds to the
tilted plane $(x^{*0},\3x^*) = (\gamma \3v\cdot\3x,\g\3x)$, but
because $\phi_\cm$ is independent of $x^{*0}$, the effect of the tilt
to the wave function vanishes and \eq\nr{philf} has the simple form
\be
  \phi_\lf(0,\3x) = \phi_\cm(\g\3x).
\la{lfcmrelation}
\ee
Equation \nr{lfcmrelation} relates the laboratory frame wave function
\be
  \psi_\lf(0,\3x,t,\3X) = e^{-iW_\lf t + i\3P\cdot\3X}\phi_\lf(0,\3x)
\la{psilf}
\ee
to the center of mass frame wave function \eq\nr{psicm}.  The total
energy of the system in the two frames is related by $W_\lf^2 =
W_\cm^2 +\3P^2$.  Finally, from \eqs\nr{pequation} and \nr{psicm} we
note that $\phi_\cm$ satisfies the Helmholtz equation
\be
  (\nabla^2_{\3x^*} + p^{*2}) \phi_\cm(\3x^*) = 0
\la{phicmeq}
\ee
where $p^{*2} = (W_\cm/2)^2 - \m^2$.  Equations \nr{lfcmrelation} and
\nr{phicmeq} will be of fundamental importance when we consider the
wave functions on a torus: the boundary conditions imposed by the
torus in the laboratory frame are transformed by \eq\nr{lfcmrelation}
into boundary conditions on the solutions \eq\nr{phicmeq}.  In what
follows we shall omit the superscript $^*$ from the center of mass
frame quantities.

\subsection{The scattering wave function\la{sec:wavefunction}}

The covariant formulation of an interacting two-particle theory is a
non-trivial task.  However, the details of the interaction are not
essential for the problem at hand.  For concreteness, we shall
describe the interaction with a potential $A_\mu(\3x)$ in the center
of mass frame, without specifying the nature of the potential or how
it acts on the wave function.  It is sufficient to assume only that
the equations that replace the Klein-Gordon equations \nr{KleinGordon}
when $A_\mu$ is included admit a square integrable solution and that
the potential has a finite range \cite{Luscher91}:
\be
  A_\mu(\3x) = 0 \h \mbox{for} \h |\3x| > R.
\la{potential}
\ee
Let us assume that there exists $R$ so that \eq\nr{potential} is true
both in the center of mass and laboratory frames.  Then the
Klein-Gordon equations
\nr{KleinGordon} hold when $|\3x| > R$, and in this region the wave
functions \eq\nr{psicm} and \eq\nr{psilf} are related by
\eq\nr{lfcmrelation}.

In the center of mass frame the interaction is spherically symmetric.
The wave function can be expanded in spherical harmonics
\be
  \phi_\cm(\3x) = \sum_{l=0}^{\infty} \sum_{m=-l}^{l}
	Y_{lm}(\theta,\varphi) \phi_{lm}(x)
\la{spherical}
\ee
where $\3x = x(\sin\theta \cos\varphi, \sin\theta \sin\varphi,
\cos\theta)$.  When $x > R$, $\phi_\cm$ is a solution of \eq\nr{phicmeq},
and the functions $\phi_{lm}$ satisfy the radial differential equation
\be
  [\frac{\dd^2}{\dd x^2} + \frac{2}{x} \frac{\dd}{\dd x}
   - \frac{l(l+1)}{x^2} + p^{2}] \phi_{lm}(x) = 0
\la{radialeq}
\ee
where $p^{2} = (W_\cm/2)^2 - \m^2 > 0$.  The solutions of
\eq\nr{radialeq} can be written as linear combinations of the
spherical Bessel functions
\be
  \phi_{lm}(x) = c_{lm} \left[ a_l(p) j_l(p x)
		+ b_l(p) n_l(p x) \right].
\la{bessel}
\ee

In the region $x < R$ the form of the radial equation for the
coefficients $\psi_{lm}$ is unknown.  We shall assume that for each
$l,m$ there exists an unique regular solution for all values of
$W_\cm>2\m$ (for a more detailed discussion, see
\cite{Luscher86,Luscher91}).  The coefficients $c_{lm}$, $a_l$ and
$b_l$ are then determined when the interior and exterior solutions
are joined together at $x = R$.

By comparing the function defined in equations \nr{spherical} and
\nr{bessel} to the definition of the phase shift $\delta_l$ given
in \eq\nr{partial-amplitude}, we can establish the well-known
connection between the scattering phase shift and the coefficients
$a_l$ and $b_l$ \cite{Landau}:
\be
   e^{i 2\delta_l(p)} =
  \fr{a_l(p) + ib_l(p)}{a_l(p) - ib_l(p)}.
\la{phaseshift}
\ee
Since $a_l$ and $b_l$ can be chosen real-valued when $p > 0$,
$\delta_l(p)$ is a real analytic function.  For a fixed $l$-sector,
the phase shift can now be given in terms of the laboratory frame
energy with the relation $p^{2} = (W_\lf^2 - \3P^2)/2 - \m^2$.

\subsection{Eigenstates on a torus\la{sec:states-on-torus}}

Let us now study the system in a (laboratory frame) box of size
$L\times L\times L$ with periodic boundary conditions.  The time
direction of the box is taken to be infinite.  The laboratory frame
wave functions $\psi_\lf$ are periodic with respect to the position of
either of the particles:
\be
  \psi_\lf(\3x_1,\3x_2) = \psi_\lf(\3x_1 + \3nL,\3x_2 + \3mL) \h\h
  \mbox{for all}\h \3n,\3m \in \Z^3\,.
\la{lfperiod}
\ee
The form of the wave function $\psi_\lf$ is given by \eq\nr{psilf}
\be
  \psi_\lf(\3x_1,\3x_2) = e^{i\3P\cdot(\3x_1+\3x_2)/2}
		\phi_\lf(\3x_1 - \3x_2).
\la{lfansatz}
\ee
Equations \nr{lfperiod} and \nr{lfansatz} together yield the result
\be
\begin{array}{rcl}
  \3P &=& (2\pi/L)\,\3d \\
  \phi_\lf(\3x) &=& (-1)^{\3d\cdot\3n} \phi_\lf(\3x+\3nL)
\end{array}
\h\h \mbox{}\h \3d,\3n \in \Z^3\,.
\la{lfquant}
\ee
The total momentum $\3P$ is conserved, and the quantization rule
\nr{lfquant} divides the wave functions into discrete total momentum
sectors, which we classify by the vector $\3d\in\Z^3$.  We are mostly
interested in sectors $\3d=(0,0,0)$ and $\3d=(0,0,1)$ (and
permutations).  Depending on the value of $\3d$, the function
$\phi_\lf$ is either periodic or antiperiodic with period length $L$.

Now we can use equation \nr{lfcmrelation} to obtain the
corresponding periodicity rule for the center of mass wave function.
For a fixed vector $\3d$,
\be
  \phi_\cm(\3x) = (-1)^{\3d\cdot\3n} \phi_\cm(\3x+\g\3nL) \h\h
  \mbox{for all} \h \3n\in \Z^3.
\la{cmperiod}
\ee
The energy $W_\lf$ and the box size $L$ determine $\g$, since
$\3v=\3P/W_\lf = 2\pi\3d/(LW_\lf)$.  Equation \nr{cmperiod} has an
simple interpretation: the center of mass system sees the laboratory
frame torus {\em expanded\,} by the fraction $\gamma$ to the direction
of the total momentum, while the length scales to the perpendicular
directions are preserved.  For compactness, we shall call the
functions obeying the rule \nr{cmperiod} {\em $\3d$-periodic\,}
functions.

Let us note that the Lorentz-contraction reduces the cubic symmetry of
the original box: if $\3d=(0,0,1)$, the volume changes to $L\times
L\times(\gamma L)$, which has tetragonal point symmetry.  If
$\3d=(1,1,0)$, the square in the $(1,2)$-plane is deformed into a
skewed parallelogram with diagonals $\sqrt{2}L$ and $\gamma\sqrt{2}L$;
only orthorhombic point symmetry remains.  In general, if all 3
components of $\3d$ are non-zero, the volume becomes a parallelepiped
with all angles between the faces non-orthogonal.  This reduction of
the symmetry is further discussed in section \ref{sec:symmetry}.

In the center of mass frame, the interaction has the same period as
the wave function but without the antiperiodicity:
\be
  A_{L,\mu}(\3x) = \sum_{\3n\in\Z^3} A_\mu (\3x+\g\3nL)\,.
\ee
Assuming that $L > 2R$, we can define the ``exterior'' region
\be
  \Omega_\cm = \{\3r\in\R^3 \Big| |\3r-\g\3nL| > R
             \mbox{~for all~} \3n\in\R^3\}
\ee
where the potential $A_L$ vanishes.  In this region $\phi_\cm$
satisfies the Helmholtz equation \nr{phicmeq}
\be
  (\nabla^2 + p^{2}) \phi_\cm(\3x) = 0\,.
\la{helmholtz}
\ee
In the region $R < r < L/2$ the solution $\phi_\cm$ of the Helmholtz
equation can be expanded in spherical harmonics and spherical Bessel
functions (see \eqs\nr{spherical} and \nr{bessel}).  Following section
\ref{sec:wavefunction}, it can be shown that there exists a unique
solution of the full interacting equations of motion in $\R^3$ which
coincides with $\phi_\cm$ in the region $\Omega_\cm$.

Our task is now to fit together the boundary condition
\eq\nr{cmperiod} and the spherical components given by
\eq\nr{spherical}.  We perform this by finding the general form
of the Helmholtz equation and expand it in spherical harmonics and
Bessel functions in the region $R < r < L/2$.

\subsection{Singular $\3d$-periodic solutions of the Helmholtz
equation\la{sec:singular}}

In this section we derive the general form of the solutions of the
Helmholtz equation obeying the periodicity rule \eq\nr{cmperiod}.
With the exception of the $\3d$-periodicity, our calculation follows
the one by L\"uscher (sections 3 and 4 in ref.~\cite{Luscher91}), and
in our discussion we shall omit technical details which can be
directly recovered by appropriate generalizations of L\"uscher's
formalism.

In the following we shall call a function $\phi$ a singular
$\3d$-periodic solution of the Helmholtz equation, when it is a smooth
function defined for all $\3x\neq\g\3nL$, $\3n\in\Z^3$, and it
satisfies the Helmholtz equation \nr{helmholtz}
\be
  (\nabla^2 + p^{2}) \phi(\3x) = 0
\ee
for some value of $p>0$, and obeys the $\3d$-periodicity rule
\be
  \phi(\3x) = (-1)^{\3d\cdot\3n} \phi(\3x + \g\3n L)
  \h\h\mbox{for all~} \3n\in\Z^3\,.
\la{dperiod}
\ee
Furthermore, we require that the function is bounded by a power of
$1/|\3x|$ near the origin:
\be
  \lim_{\3x\rightarrow 0} |\3x^{\Lambda+1} \phi(\3x) | < \infty
\ee
for some positive integer $\Lambda$, which we call the degree
of $\phi$.

In what follows we shall also assume that the value of $p$ is not
``singular'':
\be
  p\neq\fr{2\pi}{L}|\g^{-1}(\3n + \half\3d)|
  \h\h\mbox{for all~}\3n\in\Z^3.
\ee
For singular values of $p$ the Helmholtz equation admits plane wave
solutions which complicate the situation.  These values require a
separate treatment, which we shall omit here; the treatment is a
straightforward generalization of the non-singular case
\cite{Luscher91}.

We can now define the Green function
\be
  G^\3d(\3x;p) = \gamma^{-1} L^{-3} \sum_{\3k\in\Gamma}
	   \frac{e^{i\3k\cdot\3x}}{\3k^2 - p^{2}}
\la{green}
\ee
where the sum is over the momentum lattice
\be
  \Gamma = \{\3k\in\R^3 \,\Big|\, \3k = \fr{2\pi}{L}
	\g^{-1}(\3n + \half\3d)
     	\mbox{~~for some~} \3n\in\Z^3\}\,.
\ee
Since $p$ is non-singular, function \nr{green} is well-defined.  If
now $\3k = (2\pi/L)\g^{-1}(\3m+\half\3d)$ for some $\3m\in\Z^3$, then
\be
  \3k\cdot(\3x + \g\3n L) = \3k\cdot\3x + \pi\3d\cdot\3n
	+ 2\pi\3m\cdot\3n
        \h\h\mbox{for all~}\3n\in\Z^3\,,
\ee
and the function $G^\3d(\3x;p)$ is obviously $\3d$-periodic.  It
satisfies the equation
\be
  (\nabla^2 + p^{2})G^\3d(\3x;p) = - \sum_{\3n\in\Z^3}
  (-1)^{\3d\cdot\3n}  \delta(\3x+\g\3n\L)\,.
\ee
The behavior of $G^\3d$ around the origin $\3x=0$ is given by
\be
  G^\3d(\3x;p) = \fr{p}{4\pi}n_0(px) +
	(\mbox{regular part at~}\3x=0)
\ee
which follows from the fact that the Bessel function $n_0$ satisfies
the equation
\be
  (\nabla^2 + p^{2})\,n_0(p|\3x|) = -\fr{4\pi}{p}\delta(\3x)\,.
\ee
Thus, $G^\3d$ is an example of singular $\3d$-periodic solutions of
the Helmholtz equation.

Further solutions can be obtained by differentiating $G^\3d$ with
respect to $\3x$.  Let us define functions
\be
  G^\3d_{lm}(\3x;p) = \Y_{lm}(\nabla)G^\3d(\3x;p)
\ee
where we have introduced the harmonic polynomials $\Y_{lm}(\3x) =
x^lY_{lm}(\theta,\varphi)$.  Since $\Y_{lm}(\nabla)$ commutes with
$\nabla^2$, the functions $G^\3d_{lm}$ are singular $\3d$-periodic
solutions of the Helmholtz equation.  Also, it can be shown that the
functions $G^\3d_{lm}$ form a complete set of solutions in the sense
that any singular $\3d$-periodic solution of degree $\Lambda$ is a
linear combination of the functions $G^\3d_{lm}(\3x;p)$ with
$l\le\Lambda$ \cite{Luscher91}.

As discussed in section \ref{sec:wavefunction}, when $0<x<L/2$ the
functions $G^\3d_{lm}$ can be expanded in spherical harmonics.  The
expansion has the form
\be
  G^\3d_{lm}(\3x;p) =
  \fr{(-1)^l p^{l+1}}{4\pi}
  \left\{n_l(px)Y_{lm}(\theta,\varphi) +
         \sum_{l'=0}^{\infty} \sum_{m'=-l}^{l}
         M^{\3d}_{lm,l'm'}(p) \, j_{l'}(px) Y_{l'm'}(\theta,\varphi)
  \right\}
\la{Gspherical}
\ee
where the part singular at $\3x=0$ is directly computable from the
action of $\Y_{lm}(\nabla)$ to the function $n_0(px)$.  The regular
part contains coefficients $M^{\3d}_{lm,l'm'}(p)$; in practice, we
need only the first few of the coefficients, but for completeness, we
give the general expression:
\be
  M^{\3d}_{lm,l'm'}(p) = \gamma^{-1} \fr{(-1)^l}{\pi^{3/2}}
    \sum_{j=|l-l'|}^{l+l'}  \sum_{s=-j}^j \fr{i^j}{q^{j+1}}
    Z_{js}^{\3d} (1;q^2) C_{lm,js,l'm'}\,,
\la{Mexpression}
\ee
where we have defined
\be
  q = \fr{pL}{2\pi}\,.
\la{defq}
\ee
The tensor $C_{lm,js,l'm'}$ can be written in terms of Wigner
$3j$-symbols
\be
  C_{lm,js,l'm'} = (-1)^{m'} i^{l-j+l'}
	\sqrt{(2l+1)(2j+1)(2l'+1)}
       \left(
         \begin{array}{rrr}
           l & j & l' \\
           m & s & -m'
         \end{array}
       \right)
       \left(
         \begin{array}{rrr}
           l & j & l' \\
           0 & 0 & 0
         \end{array}
       \right)  \,.
\la{Cexpression}
\ee
The generalized zeta function in \eq\nr{Mexpression} is defined
through the equation
\be
  Z_{lm}^{\3d} (s; q^2) = \sum_{\3r\in P_{\rm d}}
	\Y_{lm}(\3r)(\3r^2 - q^2)^s\,,
\la{zetafunction}
\ee
where the summation is over the set
\be
  P_\3d = \{\3r\in\R^3\,|\,\3r = \g^{-1}(\3n+\half\3d)
  \mbox{~~for some~} \3n\in\Z^3\}\,.
\ee
The sum in \eq\nr{zetafunction} converges when $\mbox{Re}\,2s > l+3$,
and can be analytically continued to the whole complex plane.  The
method for evaluating the zeta function at $s=1$ is discussed in
section \ref{sec:evaluate-zeta}.

If we now choose the sector $\3d=0$, the laboratory frame and the
center of mass frame coincide, $\gamma\rightarrow 1$ and
$P_\3d\rightarrow\Z^3$, and equations
(\ref{Gspherical}--\ref{zetafunction}) reduce to the form given in
ref.~\cite{Luscher91}.  Since the derivation of the above formulae is
lengthy and analogous to the discussion therein, we shall omit it
here.

\section{Symmetry considerations\la{sec:symmetry}}

When the laboratory frame and the center of mass frame coincide, the
system exhibits a cubic symmetry and the wave functions transform
according to the representations of the cubic group O(3,$\Z$).  This
symmetry was utilized by L\"uscher \cite{Luscher91} to considerably
simplify the expressions for the energy spectrum.  However, as
mentioned in section \ref{sec:states-on-torus}, if the two frames are
not equivalent, the Lorentz boost from the laboratory frame to the
center of mass frame in effect ``deforms'' the cubical volume and only
some subgroup of the original cubic point symmetry group remains.

The deformations caused by the Lorentz boost are of a special kind:
the length scales to the direction of the boost are multiplied by
$\gamma$, whereas the perpendicular length scales are preserved.
Depending on the orientation of the boost with respect to the
directions defined by the periodicity of the (laboratory frame) torus,
we are left with different subgroups of the cubic symmetry.  First,
let us consider a boost along one of the coordinate axes, say
$\3d=(0,0,d)$.  The geometry of the box changes $(1,1,1)\rightarrow
(1,1,\gamma)$, and the relevant symmetry group is the tetragonal point
group $D_{4h}$.  This group has 16 elements: 4 rotations through an
angle $(n\pi/2)$, where $n=0,1,2,3$, around the $x_3$-axis; 4
rotations of an angle $\pi$ around directions $\3e_1$, $\3e_2$,
$\3e_1+\3e_2$ and $\3e_1-\3e_2$; and all eight of the above multiplied
by the reflection with respect to the (1,2)-plane.

When $\3d\propto(1,1,0)$, the unit square in the (1,2)-face is
deformed into a parallelogram with edges of length
$[(1+\gamma^2)/2]^{1/2}$ and with an angle $2\tan^{-1}\gamma$ between
adjacent edges.  The symmetry group is the orthorhombic group $D_{2h}$
with 8 elements.

The relevant point groups and the boost vectors are classified in
table \ref{table:symmetry}.  All of the groups contain the parity
transformation $\3x\rightarrow -\3x$.  The group $C_i$ has only two
elements, the identity element and the parity transformation; it
results from a boost where the length of all of the 3 components of
$\3d$ are different from each other.

\begin{table}[tb]
\centering
\begin{tabular}{cclc} \hline
 $\3d$     &  point group & classification &
	 $N_{\rm elements}$ \\ \hline
 $(0,0,0)$ & $O_h$    & cubic          & 48  \\
 $(0,0,a)$ & $D_{4h}$ & tetragonal     & 16  \\
 $(0,a,a)$ & $D_{2h}$ & orthorhombic   &\n8  \\
 $(0,a,b)$ & $C_{2h}$ & monoclinic     &\n4  \\
 $(a,a,a)$ & $D_{3d}$ & trigonal       & 12  \\
 $(a,a,b)$ & $C_{2h}$ & monoclinic     &\n4  \\
 $(a,b,c)$ & $C_i$    & triclinic      &\n2  \\  \hline
\end{tabular}
\caption[1]{The classification of the Lorentz boosts on a torus and
the reduction of the cubic symmetry.  The first column displays the
direction of the boost (modulo permutations); the numbers $a$, $b$ and
$c$ are all taken to be different from each other and from 0.  The
notation used for the groups is the Schonflies notation.
\cite{Weissbluth}\la{table:symmetry}}
\end{table}

In the simulations described in this paper we use only the two lowest
total momentum sectors, $|\3d| = 0$ or 1.  The next two total momentum
sectors suffer from the fact that already the two lowest energy levels
are nearly degenerate (see section \ref{sec:noninteract}), which
limits the usability of these sectors in practical Monte Carlo
simulations.  Thus, in the following we discuss mainly only the cubic
and tetragonal symmetry groups $O_h$ and $D_{4h}$.  We shall
explicitly select vectors $\3d=\30$ (cubic symmetry, $O_h$) and
$\3d=(0,0,1)$ (tetragonal symmetry, $D_{4h}$).  This choice gives the
$x_3$-axis a special status, and many calculations are simpler to
perform; naturally, the final results are independent of the choice of
the axis.  At this stage we need not invoke the antiperiodicity of the
wave functions to the $x_3$-direction, and the formulae given below
are valid for any $\3d=(0,0,n)$.

The symmetry properties of the zeta function $Z_{lm}^\3d$ follow
directly from the definition \eq\nr{zetafunction} and the properties
of the spherical harmonic functions under symmetry operations.  In
particular, let us consider the following transformations: parity
transformation $\3x\rightarrow -\3x$, a reflection
$\3x\rightarrow(x_1,-x_2,x_3)$, and a rotation of an angle $\pi/2$
around $x_3$-axis.  These operations yield
\be
\begin{array}{rcl}
  Z_{lm}^{\3d}(s;q^2) &=& 0  \h\mbox{if $l$ is odd}   \\
  Z_{lm}^{\3d}(s;q^2) &=& Z_{l-m}^{\3d}(s;q^2)        \\
  Z_{lm}^{\3d}(s;q^2) &=& 0  \h\mbox{if $m\neq 4n$, $n\in\Z$}  \\
\end{array}
\la{z-symmetries}
\ee
The first property is valid for all of the groups listed in
table~\ref{table:symmetry}.  The cubic group has additional symmetries
absent from the octahedral group.  Most important of these are
$Z^0_{20} = 0$ and $Z^0_{44} = \sqrt{70}/14\,Z^0_{40}$.  These
differences have an effect in the energy spectrum calculation.

Using the above properties of the zeta function and symmetries of the
3$j$-symbols \cite{Weissbluth,Rotenberg} we can derive symmetry
relations for the matrix elements $M^\3d_{lm,l'm'}$ defined in
\eq\nr{Mexpression}:
\be
\begin{array}{l}
  M^\3d_{lm,l'm'} = 0 \h\h \mbox{if~} l'\neq l \pmod2    \\
  M^\3d_{lm,l'm'} = 0 \h\h \mbox{if~} m'\neq m \pmod4    \\
  M^\3d_{lm,l'm'} = M_{l'm',lm}^\3d = M_{l(-m),l'(-m')}^\3d \,.
\end{array}
\la{M-symmetries}
\ee
In table \ref{table:Mtable} we list the expressions of
$M_{lm,l'm'}^\3d$ for $l,l'\le 3$.  In order to simplify the table, we
have defined
\be
  w_{lm} = \fr{1}{\pi^{3/2}\sqrt{2l+1}}
	\gamma^{-1}q^{-l-1}\, Z_{lm}^\3d(1;q^2)\,.
\la{wlm}
\ee
The necessary $3j$-symbol values can be found in the literature
\cite{Rotenberg}.  Matrix elements missing from the table are either
zero, or can be obtained with the symmetry
relations~\nr{M-symmetries}.

\begin{table}
\setlength{\templength}{\tabcolsep}
\setlength{\tabcolsep}{0.5mm}
\centering
\begin{tabular}{rrrrr|rrrrrr}\hline
$l$ & $m$ & $l'$ & $m'$ &\n&
	\multicolumn{6}{c}{$M_{lm,l'm'}^\3d$} \\
\hline\hline
 0&\n\n0&\n\n0&\n\n0&&
	     \n\n$w_{00}$&$                     $&$
                        $&$                     $&$
                        $&$                     $ \\
\hline
 1 & 0 & 1 & 0 &&$w_{00}$&$           + 2 w_{20}$&$
                        $&$                     $&$
                        $&$                     $ \\
 1 & 1 & 1 & 1 &&$w_{00}$&$             - w_{20}$&$
                        $&$                     $&$
                        $&$                     $ \\
\hline
 2 & 0 & 0 & 0 &&$      $&$     -\sqrt{5} w_{20}$&$
                        $&$                     $&$
                        $&$                     $ \\
 2 & 0 & 2 & 0 &&$w_{00}$&$  + \fr{10}{7} w_{20}$&$
     + \fr{18}{7} w_{40}$&$                     $&$
                        $&$                     $ \\
 2 & 1 & 2 & 1 &&$w_{00}$&$      +  \fr57 w_{20}$&$
     - \fr{12}{7} w_{40}$&$                     $&$
                        $&$                     $ \\
 2 & 2 & 2&$-2$&&$      $&$                     $&$
                        $&$ \fr37\sqrt{70}w_{44}$&$
                        $&$                     $ \\
 2 & 2 & 2 & 2 &&$w_{00}$&$  - \fr{10}{7} w_{20}$&$
          + \fr37 w_{40}$&$                     $&$
                        $&$                     $ \\
\hline
 3 & 0 & 1 & 0 &&$      $&$-\fr37\sqrt{21}w_{20}$&$
   -\fr47\sqrt{21}w_{40}$&$                     $&$
                        $&$                     $ \\
 3 & 1 & 1 & 1 &&$      $&$-\fr37\sqrt{14}w_{20}$&$
   +\fr37\sqrt{14}w_{40}$&$                     $&$
                        $&$                     $ \\
 3 & 3 & 1&$-1$&&$      $&$                     $&$
                        $&$       2\sqrt3 w_{44}$&$
                        $&$                     $ \\
 3 & 0 & 3 & 0 &&$w_{00}$&$       + \fr43 w_{20}$&$
    + \fr{18}{11} w_{40}$&$                     $&$
    +\fr{100}{33} w_{60}$&$                     $ \\
 3 & 1 & 3 & 1 &&$w_{00}$&$             + w_{20}$&$
    +  \fr{3}{11} w_{40}$&$                     $&$
    - \fr{25}{11} w_{60}$&$                     $ \\
 3 & 2 & 3&$-2$&&$      $&$                     $&$
                        $&$ \fr{3}{11}\sqrt{70} w_{44}$&$
                        $&$+\fr{10}{11}\sqrt{14}w_{64}$ \\
 3 & 2 & 3 & 2 &&$w_{00}$&$                     $&$
     -\fr{21}{11} w_{40}$&$                     $&$
     +\fr{10}{11} w_{60}$&$                     $ \\
 3 & 3 & 3&$-1$&&$      $&$                     $&$
                        $&$ \fr{3}{11}\sqrt{42} w_{44}$&$
                        $&$-\fr{5}{33}\sqrt{210}w_{64}$ \\
 3 & 3 & 3 & 3 &&$w_{00}$&$       - \fr53 w_{20}$&$
      +\fr{9}{11} w_{40}$&$                     $&$
      -\fr{5}{33} w_{60}$&$                     $ \\
 \hline\hline
\end{tabular}
\caption[1]{Matrix elements $M_{lm,l'm'}^\3d$ for $\3d=(0,0,d)$ and
for $l,l' \le 3$.\la{table:Mtable}}
\setlength{\tabcolsep}{\templength}
\end{table}

The table~\ref{table:Mtable} can be compared to table E.1 in
ref.~\cite{Luscher91}, which lists the matrix elements for $\3d=0$.
The main difference is the appearance of functions $w_{20}$, $w_{44}$
and $w_{64}$ in table \ref{table:Mtable}.  If we set $\3d=0$, then
$w_{20}\rightarrow 0$, $w_{44}\rightarrow\sqrt{70}/14 w_{40}$ and
$w_{64}\rightarrow -\sqrt{14}/2 w_{60}$, and L\"uscher's result is
recovered.

If we further reduce the symmetry of the system, more of the elements
become non-zero.  For example, in the case of the group $D_{2h}$ the
element $w_{22}\ne 0$; and finally, for a generic boost vector $\3d$
the relevant group is $C_i$ and all $w_{(2l)m}\ne 0$.

\subsection{Energy spectrum\la{sec:energy}}

We are now in a position to consider the singular $\3d$-periodic
solutions of the equations of motion that contain the 2-particle
interaction.  The general form of the solutions of the equations of
motion in the region $R<|\3x|< L/2$ was given in
equations~\nr{spherical} and~\nr{bessel}.  On the other hand, in
section~\ref{sec:singular} it was noted that the functions
$G_{lm}^{\3d}(\3x,p^2)$ form a complete set of singular $\3d$-periodic
solutions when $l\le\Lambda$, where $\Lambda$ is the degree of the
function.  When we demand that the functions are equal, we have
\be
  \sum_{l=0}^\Lambda \sum_{m=-l}^{l} v_{lm} G_{lm}^{\3d} (\3x,p^2) =
  \sum_{l=0}^\Lambda \sum_{m=-l}^{l} c_{lm}[a_l(p)j_l(px) +
  b_l(p)n_l(px)]\,Y_{lm}(\theta,\varphi)
\ee
for some constants $c_{lm}$ and $v_{lm}$.  Using equation
\nr{Gspherical}, we can eliminate $v_{lm}$ and obtain
\be
  c_{lm}a_l(p) =  \sum_{l'=0}^\Lambda \sum_{m'=-l'}^{l'}
	c_{l'm'} b_{l'}(p) M_{l'm',lm}^{\3d}(p)\,.
\la{c-eq}
\ee
A non-zero solution for the vector $c_{lm}$ exists if the determinant
of a matrix we are about to construct vanishes.  We rewrite
\eq\nr{c-eq} as a matrix equation $C(A-BM) = 0$, where
matrix $A_{(lm),(l'm')} = a_l(p) \delta_{l,l'}\delta_{m,m'}$ (likewise
for $B$).  Since $A$ and $B$ are diagonal and all the diagonal
elements of $A-iB$ are non-zero, we can utilize relation
\nr{phaseshift} to define the phase shift matrix
\be
  e^{2i\delta} = (A+iB)/(A-iB)
\ee
Noting that $A-BM = (i/2)(A+iB)(M-i) - (A-iB)(M+i)$, we can factor out
$(A-iB)$ and the determinant condition acquires the compact form
\be
  \det \left[e^{2i\delta} (M-i) - (M+i)\right] = 0\,.
\la{determinant}
\ee
This equation is equivalent to eq.~(4.10) in ref.~\cite{Luscher91}.

Using the symmetry properties \nr{M-symmetries} of the matrix elements
$M_{lm,l'm'}^{\3d}$, it is evident that the sectors with even and odd
$l$ in \eq\nr{c-eq} are completely independent of each other.  This is
a direct consequence of the symmetry of the system under reflections
$\3x\rightarrow-\3x$: reflection symmetry implies that the energy
eigenstates are also eigenstates of parity, and the the spherical
harmonics have parity $P = (-1)^l$.

Generally, the energy eigenstates will belong to some irreducible
representation of the symmetry group of the system.  The cubic group
of proper rotations $O$ has 5 representations: 2 dimension 1
representations $A_1$ and $A_2$, 2-dimensional representation $E$, and
3-dimensional representations $T_1$ and $T_2$.  $O_h$ includes the
parity transformation and the number of representations is doubled to
parity even and odd representations, which are denoted with $A_1^+$
and $A_1^-$ (likewise for the other representations).  The tetragonal
group $D_4$ has 4 1-dimensional representations $A_1$, $A_2$, $B_1$,
$B_2$, and one 2-dimensional representation $E$ \cite{Weissbluth}.
Again, the group $D_{4h}$ doubles these to parity even (${}^+$) and
odd (${}^-$) representations.

The cubic symmetry has been covered in detail in \cite{Luscher91}, and
we concentrate here on the tetragonal symmetry.  The representations
$\Gamma^{(j)}$ of $O(3)$ with $l\le 3$ are reduced into irreducible
representations of $D_{4h}$ as follows:
\be
\begin{array}{rcl}
  \Gamma^{(0)} &=& A_1^+   \\
  \Gamma^{(1)} &=& A_2^- \oplus E^- \\
  \Gamma^{(2)} &=& A_1^+ \oplus B_1^+ \oplus B_2^+ \oplus E^+ \\
  \Gamma^{(3)} &=& A_2^- \oplus B_1^- \oplus B_2^- \oplus 2\,E^-
\end{array}
\la{reduce}
\ee
The representations can be identified by using character tables
\cite{Weissbluth} or by enumerating harmonic polynomials of degree $l$
which transform according to the representations of $D_{4h}$.  The
basis polynomials for each of the representations are listed in
table~\ref{table:basis} for $l\le 3$; the polynomials are linear
combinations of the harmonic polynomials $\Y_{lm}(\3x)$ for each
$l$-sector.  The representation $E^-$ occurs twice in the reduction of
$\Gamma^{(3)}$ and has two sets of basis polynomials.

\begin{table}[tb]
\[\begin{array}{c|llllrcl} \hline
\mbox{representation}
      &  l=0\n&  l=1\n &  l=2                  &  l=3  &
\multicolumn{3}{c}{\mbox{indices}} \\
\hline
A_1^+ &1&   &x_3^3 - \fr13x^2\n    &                     && & \\
A_2^- & &x_3&                      & x_3^3 - \fr35 x^2x_3&& & \\
B_1^+ & &   &x_1^2 - x_2^2         &                     && & \\
B_2^+ & &   &x_1x_2                &                     && & \\
B_1^- & &   &                      & x_1x_2x_3           && & \\
B_2^- & &   &                      & (x_1^2 - x_2^2)x_3  && & \\
E^+   & &   &x_ix_3                &     &  i&=&1,2           \\
E^-   & &x_i&    & (x_i^2 - \fr35x^2)x_i &  i&=&1,2           \\
      & &   &    & (x_i^2 - x_3^2)x_j    & (i,j)&=&(1,2),(2,1)\\
\hline
\end{array}\]
\caption[1]{The basis polynomials of the irreducible representations
of $D_{4h}$.\la{table:basis}}
\end{table}

The fact that the eigenstates belong to the irreducible
representations of the tetragonal group becomes evident when we study
closer the matrix elements $M^{\3d}_{lm,l'm'}$: let the angular
momentum cutoff $\Lambda = 3$, and let us look at the even $l$-sector
of the matrix $M$ (the odd $l$-values are decoupled in any case).
{}From table \ref{table:Mtable} we observe that $M$ has a block diagonal
form ($m_{lm,l'm'} = M^\3d_{lm,l'm'}$)
\be
  M = \left(
  \begin{array}{cccccc}
  m_{00,00} & m_{00,20} &   &       &        &    \\
  m_{20,00} & m_{20,20} &   &       &        &    \\
            &   & m_{2-1,2-1} &     &        &    \\
            &   &    & m_{21,21}    &        &    \\
            &   &    &   &  m_{2-2,2-2} & m_{2-2,22} \\
            &   &    &   &  m_{22,2-2}  & m_{22,22}
  \end{array}
  \right)
\ee
The upper left $2\times 2$ matrix transforms according to the
representation $A_1^+$, the two equal elements at the center comprise
the two components of the 2-dimensional representation $E^+$, and if
we diagonalize the lower right $2\times 2$ matrix and symmetrize with
equations \nr{M-symmetries}, we obtain diagonal elements
$\half(m_{22,22}+m_{2-2,2-2})\pm\half(m_{22,2-2}+m_{2-2,22})$ which
operate on $B_1^+$ and $B_2^+$, respectively.

The sector $l$ odd decomposes in an analogous way.  In the following
we study the sectors $A_1^+$ and $A_2^-$, $E^-$, which are relevant
for the scalar and vector scattering channels in infinite volume.

\subsubsection{$A_1^+$ -sector}

First, let us consider the case where the angular momentum cutoff
$\Lambda=0$ or 1.  {}From the reduction \eq\nr{reduce} and
table~\ref{table:basis} we see that only $M^\3d_{00,00}$ belongs to
this sector, and \eq\nr{determinant} is one-dimensional.  It can be
written to the form
\be
  \tan\delta_0(p) = \fr{1}{M^\3d_{00,00}} =
  \fr{\gamma q \pi^{3/2}}{Z^\3d_{00}(1;q^2)} \h\h q = \fr{L}{2\pi}p
\la{det-1dim}
\ee
where \eq\nr{wlm} was used in the second equality.  This is the
fundamental result given in \eqs(\ref{deltal}--\ref{phil}), and used
in the simulations described in section \ref{sec:montecarlo}.

If $\Lambda=2$ or 3, then the sector $l=2$ is included, and the matrix
in \eq\nr{determinant} is 2-dimensional.  The determinant condition
then mixes together phase shifts $\delta_0$ and $\delta_2$,
corresponding to the infinite volume $l=0$ scalar and $l=2$ tensor
scattering channels:
\be
\begin{array}{r}
  [e^{2i\delta_0}(m_{00}-i) - (m_{00}+i)]
  [e^{2i\delta_2}(m_{22}-i) - (m_{22}+i)]\,\, \\
  =m_{20}^2 (e^{2i\delta_0} - 1)(e^{2i\delta_2} - 1)\,,
\end{array}
\la{det-2dim}
\ee
where we defined $m_{ab}\equiv M^\3d_{a0,b0}$.  If $\delta_2=0$ (mod
$\pi$), \eq\nr{det-2dim} reduces immediately to equation
\nr{det-1dim}, as expected.  Let us now consider the case $\delta_2\ne
0$.  Usually it is reasonable to assume that the low energy scattering
amplitude is dominated by the lowest $l$-channel and that the phase
shifts at higher $l$:s are small.  If we expand $\delta_0=\delta_0^0
+\Delta_0$, where $\delta_0^0$ satisfies
\eq\nr{det-1dim}, the first order correction due to \eq\nr{det-2dim}
is
\be
  \Delta_0(p) = -\fr{m_{20}^2}{m_{00}^2 + 1} \delta_2(p)\,.
\la{Delta0}
\ee
The function $m_{20}^2/(m_{00}^2+1)$ is not naturally
small: for example, for the data in case C in section
\ref{sec:montecarlo} it varies in the range 0.1 -- 14\@.  When
$p\rightarrow 0$, it diverges as $p^{-4}$ (for fixed $\gamma$ and
$L$).  However, this is not a problem, since it is offset by
$\delta_2(p)\sim p^5$ behavior at small $p$.  Nevertheless, there is
no ``built-in'' mechanism which would automatically render the
coupling between the $l=2$ channel and the $l=0$ channel small.  In
order for the \eq\nr{det-1dim} to be a good approximation, the phase
shift $\delta_2(p)$ has to be small due to the physics of the
interacting particles.  Luckily, the case is usually so: the
scattering of two particles is dominated by the lowest allowed angular
momentum channel.  For the model used in section
\ref{sec:montecarlo}, the good fit of the data to the perturbative
ansatz supports this view.

It should be noted that if $\delta_2(p)$ is not small, it becomes
impossible to extract the phase shift functions from the energy
spectrum: there are two unknown functions $\delta_0(p)$ and
$\delta_2(p)$ but only one equation \nr{det-2dim}.  Naturally, if the
phase shifts are known, it is still possible to calculate the energy
spectrum.

In the $\3d=0$ sector the matrix element $M^\3d_{20,00}$ vanishes and
equation \nr{det-2dim} reduces to equation \nr{det-1dim} and to a
similar equation for $\delta_2(p)$.  The lowest angular momentum
channel mixing with the scalar $l=0$ channel is $l=4$.  In this case
one can derive expressions similar to \eqs\nr{det-2dim} and
\nr{Delta0}, with the substitution $l=2\rightarrow 4$
\cite{Luscher91}.

\subsubsection{$A_2^-$ and $E^-$ -sectors}

$A_2^-$ and $E^-$ -sectors are important for scattering through the
$l=1$ vector channel in the infinite volume; the most famous example
being the $\pi\pi\leftrightarrow\rho$ -scattering in QCD.  Let us
assume that the angular momentum cutoff $\Lambda\le 3$.  If the cubic
symmetry were unbroken, $O(3)$ representation $\Gamma^{(1)}$ would
correspond to single 3 dimensional cubic representation $T^-$.  The
situation is now complicated by the appearance of two tetragonal
representations already at the lowest angular momentum level.  The
breaking of the cubic symmetry implies that each eigenstate splits
into two states with non-degenerate energies.

When the energy spectrum of the system is measured, one has to know to
which of the two representations the state corresponding to the energy
level belongs.  The phase shift $\delta_1(p)$ is obtained from the
determinant condition~\nr{determinant}
\be
  \tan\delta_1(p) =  \fr{1}{M^\3d(\Gamma_{D_{4h}})}
\la{delta1}
\ee
where, using table~\ref{table:Mtable},
\ba
  M^\3d(A_2^-) &=& \fr{q^{-1}}{\gamma\pi^{3/2}}
\Big[Z^\3d_{00}(1;q^2) +
\fr{2}{\sqrt{5}}q^{-2} Z^\3d_{20}(1;q^2)\Big]
	\la{ma2-}\\
  M^\3d(E^-)   &=& \fr{q^{-1}}{\gamma\pi^{3/2}}
\Big[Z^\3d_{00}(1;q^2) -
\fr{1}{\sqrt{5}}q^{-2} Z^\3d_{20}(1;q^2)\Big]\,.
	\la{me-}
\ea
It is possible to construct operators which transform according to the
representations $A_2^-$ and $E^-$.  Operators $O(A_2^-)$ decouple from
eigenstates which belong to the representation $E^-$ (and vice versa).
The correlation function matrix constructed from operators $O(A_2^-)$
gives energy eigenvalues which can be used to calculate the phase
shift through \eq\nr{ma2-}; similarly \eq\nr{me-} is used for the
eigenvalues from $O(E^-)$ correlation matrix.

If we set $d\rightarrow 0$, then $Z^\3d_{20}\rightarrow 0$,
$\gamma\rightarrow 1$, and equations \nr{ma2-} and \nr{me-} become
degenerate.  Equation \nr{delta1} reduces to form similar to the
$\delta_0(p)$ equation \nr{det-1dim} \cite{Luscher91}.

\subsection{Evaluation of the zeta function\la{sec:evaluate-zeta}}

The method for evaluating the zeta function when $\3d=0$ has been
described in detail by M.~L\"uscher \cite{Luscher91}.  The formalism
used there is easily adaptable to the $\3d\ne 0$ -case, and here we
shall give the necessary formulae for numerically evaluating the zeta
function without derivation.

Let us define a {\em heat kernel\,} of the Laplace operator on a
$\3d$-periodic torus of size $(2\pi)^3$ (so that the periodicity is
given by $\phi(\3x) = (-1)^{\3d\cdot\3n}\phi(\3x+2\pi\g\3n)$,
cf.~equation \nr{cmperiod})
\ba
  K^\3d_{lm}(t,\3x)
  &=& (2\pi)^{-3/2} \sum_{\3r\in P_d}
	\Y_{lm}(\3r) \exp(i\3r\cdot\3x - t\3r^2) \la{kernel1} \\
  &=& \gamma (4\pi t)^{-3/2} \fr{i^l}{(2t)^l} \sum_{\3n\in\Z^3}
	(-1)^{\3d\cdot\3n} \Y_{lm}(\3x + 2\pi\g\3n)
	\exp[-\fr{1}{4t} (\3x + 2\pi\g\3n)^2]      \la{kernel2}
\ea
where
$P_\3d=\{\3r\in\Z^3\,|\,\3r=\g^{-1}(\3n+\half\3d),\n\3n\in\Z^3\}$.
The first expression is useful when $t$ is large, and the second when
$t$ is small.  Defining the truncated kernel
\be
  K_{lm}^{\3d,\lambda}(t,\3x) = K^\3d_{lm}(t,\3x) - \sum_{
		\stackrel{\scriptstyle \3r \in P_d}{|\3r| < \lambda}}
	 \Y_{lm}(\3r) \exp(i\3r\cdot\3x - t\3r^2)
\ee
it can be shown that the zeta function has a rapidly convergent
integral expression
\be
  Z^\3d_{lm}(1;q^2) =
  \sum_{\stackrel{\scriptstyle \3r \in P_d}{|\3r| < \lambda}}
  \fr{\Y_{lm}(\3r)}{\3r^2 - q^2}
  + (2\pi)^3 \int_0^\infty \dd t \left( e^{tq^2}
  K_{lm}^{\3d,\lambda}(t,\30) -
  \fr{\gamma\delta_{l,0}\delta_{m,0}}{(4\pi)^2 t^{3/2}}\right).
\ee
In the integral above one uses the kernel expression \nr{kernel2} when
$t<1$, and \nr{kernel1} otherwise.  The cutoff $\lambda$ is chosen so
that $\lambda^2 > \mbox{Re\,} q^2$.

\section{Conclusions\la{sec:conclusions}}

In this paper, we have developed a method that enables one to measure
the elastic scattering phase shift with lattice Monte Carlo
simulations when the total momentum of the scattering particles is
non-zero.  The method is an extension of the zero total momentum
formalism, originally developed by
L\"uscher~\cite{Luscher86,Luscher91}, and it is based on the energy
spectrum of the two-particle states in finite volumes.

We have applied the method to a 4-dimensional test model consisting of
two Ising spin fields.  By suitably choosing the parameters, the
fields correspond to particles of two different masses, and by
introducing a coupling between the fields the heavy particle appears
as a resonance in the S-wave scattering channel of two light
particles.  We have shown excellent consistency between the rest frame
and non-rest frame results and also the perturbation theory, and we
have been able to extract the properties of the resonance to a good
accuracy.

In our Monte Carlo simulations we have used a large number of lattice
sizes and correspondingly a wide range of lattice momenta.  This has
enabled us to obtain a good description of the phase shift function
for the whole momentum range of interest separately for both the rest
frame and the non-rest frame scattering sectors.  However, often one
has access only to a limited range of lattice sizes.  Then the two
momentum sectors complement each other: for a given lattice size, the
center of mass frame energy and the relative momentum of the
scattering particles is different in the total momentum $\3P=0$ and
$\3P=2\pi/L$ sectors.  Moreover, due to the kinematics of the
scattering particles, the resonance usually appears at a smaller
lattice size in the non-zero momentum sector than in the zero momentum
sector.  This can be a very significant factor when the lattice model
is computationally so demanding that simulations in very large volumes
are impractical.

As a final note, we consider the success of our work in this paper to
be quite encouraging: we have obtained high-precision data for
scattering phases in a 4-dimensional quantum field theory with only
workstation-class computer simulations.  It would be both interesting
and challenging to apply these techniques to QCD.

\subsubsection*{Acknowledgments}
We would like to thank J.~Westphalen for the initial version of the
code to evaluate the phase shift function.  We are grateful to
the U.~S.~Dept. of Energy for their support under
grant No. DE-FG02-91ER40661, and to
Indiana University Computing Services for their
support of the computations.

\end{document}